\documentclass[twocolumn,showpacs,showkeys,preprintnumbers,amssymb,amsmath,aps,superscriptaddress,prb]{revtex4-1}
\usepackage{graphicx}
\usepackage{dcolumn}
\usepackage{bm}
\usepackage{braket}
\usepackage{color}
\usepackage{mathtools}

\begin{document}


\title{Non-perturbative linked-cluster expansions in long-range ordered quantum systems}

\author{Dominik Ixert}
\email{dominik.ixert@tu-dortmund.de}
\affiliation{Lehrstuhl f\"ur Theoretische Physik I, Otto-Hahn-Strasse 4, TU Dortmund, D-44221 Dortmund, Germany}

\author{Kai Phillip Schmidt}
\email{kai.phillip.schmidt@fau.de}
\affiliation{Lehrstuhl f\"ur Theoretische Physik I, Staudtstra{\ss}e 7, Universit\"at Erlangen-N\"urnberg, D-91058 Erlangen, Germany}

\date{\today}

\begin{abstract}
We introduce a generic scheme to perform non-perturbative linked cluster expansions in long-range ordered quantum phases. Clusters are considered to be surrounded by an ordered reference state leading to effective edge-fields in the exact diagonalization on clusters which break the associated symmetry of the ordered phase. Two approaches, based either on a self-consistent solution of the order parameter or on minimal sensitivity with respect to the ground-state energy per site, are formulated to find the optimal edge-field in each NLCE order. Furthermore, we investigate the scaling behavior of the NLCE data sequences towards the infinite-order limit. We apply our scheme to gapped and gapless ordered phases of XXZ Heisenberg models on various lattices and for spins 1/2 and 1 using several types of cluster expansions ranging from a full-graph decomposition, rectangular clusters, up to more symmetric square clusters. It is found that the inclusion of edge-fields allows to regularize non-perturbative linked-cluster expansions in ordered phases yielding convergent data sequences. This includes the long-range spin-ordered ground state of the spin-1/2 and spin-1 Heisenberg model on the square and triangular lattice as well as the trimerized valence bond crystal of the spin-1 Heisenberg model on the kagome lattice. 
\end{abstract}


\maketitle

\section{Introduction}
The search for exotic phases of correlated quantum systems have become one of the most promising -- but also very challenging -- topic within condensed matter theory. It is believed that the interplay of strong quantum fluctuations and frustrated interactions plays the most important role in stabilizing such unconvential phases. However, the most common quantum phases in condensed matter physics display spontaneous symmetry breaking and long-range order. Exotic phases are then often close in parameter space to these ordered phases and it is a notable challenge and an important task to develop strategies and tools allowing to extract quantitative information of such more conventional phases in order to characterize real systems and to guide experiments toward more exotic phases.  

A perfect microscopic test bed for this question is the antiferromagnetic Heisenberg model at zero temperature in two dimensions, since it is expected to display various ordered and disordered ground states depending on the degree of geometric frustration and on the value of the total spin.  At the same time it is the relevant microscopic description for many experiments. Important examples are the unfrustrated  spin-1/2 Heisenberg model on the square lattice which shows long-range N\'{e}el order~\cite{Manousakis91,Reger88,Chakravarty89} as well as the same model on the geometrically frustrated triangular lattice which has been studied extensively over the last decades and exhibits a three-sublattice $120^\circ$ ordered ground state~\cite{Bernu92,Capriotti99,White07}. In both cases the SU(2) symmetry is spontaneously broken and one has gapless spin-wave exitations according to the Goldstone Theorem~\cite{Auerbach94}. In contrast, on the highly frustrated kagome lattice, the spin-1/2 Heisenberg model is believed to realize a quantum spin liquid ground state with exotic topological order \cite{Yan11,Depenbrock12} while the ground state of the spin-$1$ cousin is most likely a spontaneous trimerized phase with long-range singlet order~\cite{Changlani15,Liu15,Li15,Ghosh15,Oitmaa15,Ixert15} and gapped excitations. For larger spins $S>1$, the ground-state of the Heisenberg model on the kagome lattice is again magnetically ordered~\cite{Huse92,Henley09,Goetze11,Oitmaa15}. 

Numerically, there are several techniques which can be applied to two-dimensional quantum many-body systems which all have strengths as well as complications. One promising tool which is under active development in recent years are so-called non-perturbative linked-cluster expansions (NLCEs)\cite{Rigol06,Rigol07_1,Rigol07_2,yang11,Tang13,coester15}, which are non-perturbative variants of perturbative linked-cluster expansions (LCEs) where high-order series expansions (SE) are derived in the thermodynamic limit using a full-graph decomposition and the linked-cluster theorem. In \mbox{NLCEs}, perturbation theory on graphs is replaced by non-perturbative tools like exact diagonalization (ED), density matrix renormalization group~\cite{stoudenmire14} or continuous unitary transformations \cite{yang11,coester15}. 

All current NLCEs are real-space approaches. As a consequence, one expects convergence in gapped quantum phases due to the finite correlation length and complications for gapless systems. However, as shown recently \cite{Ixert15}, even in gapped phases like the trimerized phase of the spin-$1$ Heisenberg model on the kagome lattice, the NLCE data sequences can become erratic due to peculiar quantum criticalities of lower-dimensional subsets of graphs. Here the trimerized order of the two-dimensional system is not reflected in the calculation on one-dimensional graphs leading to the wrong assignment of fluctuations which are not present in the two-dimensional system. It is therefore necessary to extend the NLCEs in such a way that each cluster ``remembers'' the quantum order out of which it is taken from. For the ordered phase of the transverse-field Ising model on the square lattice this has been already realized successfully in Ref.~\onlinecite{Kallin13}.      

In this work we present an extended scheme which also realizes this line of reasoning. We consider clusters to be taken out of a long-range ordered reference state. This state can either be a classical spin-ordered configuration like the N\'{e}el state or a valence bond solid breaking translational symmetry like a trimerized state. These reference states introduce symmetry-breaking edge-fields in the ED of clusters which enable us to perform NLCEs in gapped and gapless ordered quantum phases. We test our scheme on various lattices using either quantum spins 1/2 or 1.

This paper is organized as follows: In Sect.~\ref{Sect:nlce} we explain briefly the main idea behind NLCEs. The edge-field NLCE for ordered quantum systems is introduced in Sect.~\ref{Sect:models} and the tested scaling behaviors of NLCE data is given in Sect.~\ref{Sect:scaling}, followed by the results for the spin-$1/2$ and spin-$1$ Heisenberg models on the square and triangular lattice in Sections~\ref{Sect:results:squareOneHalf} to~\ref{Sect:results:triangularOne} displaying long-range magnetic order. Finally, in Sect.~\ref{Sect:spin1kagome}, the edge-field approach is applied to the trimerized and gapped spin-$1$ kagome Heisenberg model. The paper concludes with a brief summary of the main results.

\section{NLCE}
\label{Sect:nlce}
The essential idea of NLCEs is to exploit the linked-cluster theorem so that actual 
numerical calculations are done on finite linked clusters, but the final results are valid 
directly in the thermodynamic limit. Generically, NLCEs consist of three steps: i) choosing and generating the families of clusters or topologically distinct graphs used in the LCE, ii) performing numerical calculations on graphs extracting the physical quantities of interest, and iii) determining the reduced contributions specific to each graph and embed these contributions into the infinite lattice.

The details of the choice of graphs (i) will be given below. Concerning ii), we are using ED with the Lanczos algorithm~\cite{Lanczos50} to determine the ground-state energy $E^{\mathcal{G}_\nu}_0$ (as well as the sublattice magnetization for spin-ordered phases) on each graph. Apart from the exponential increase of graphs with the number of sites $N$, the memory needed for ED is the limiting factor of NLCEs.

In the third step iii) we concentrate on the ground-state energy per site $e_0$ and the appropriate order parameter in the thermodynamic limit, e.g.~the sublattice magnetization in long-range spin-ordered phases. In this technical part we only discuss $e_0$ which can be expressed as
\begin{equation}
e_0  = \sum_{j} \nu^{\mathcal{G}_j} \, e_{0,{\rm red}}^{\,\mathcal{G}_j} \quad , 
\label{eq:e0}
\end{equation}
where the sum runs over all linked graphs $\mathcal{G}_j$. The integer number $\nu^{\mathcal{G}_j}$ is the so-called embedding factor specifying the combinatorical number how often graph $\mathcal{G}_j$ can be embedded into the infinite lattice. The reduced contribution $e_{0,{\rm red}}^{\,\mathcal{G}_j}$ specific to graph $\mathcal{G}_j$ results from subtracting all contributions of subgraphs in order to avoid double counting. One has
\begin{equation}
 e_{0,{\rm red}}^{\,\mathcal{G}_j} = E_0 ^{\mathcal{G}_j} - \sum_{\mathcal{G}^\prime_{j}\subset \mathcal{G}_{j}} e_{0,{\rm red}}^{\,\mathcal{G}^\prime_{j}} \quad . 
\label{eq:e0_tilde}
\end{equation}
Note thate the sum runs over all connected subgraphs of $\mathcal{G}_j$ without identifying topologically equivalent contributions. Depending on the lattice, different resummation schemes could be useful. Here we apply the following four different expansions:

\subsection{Full-graph expansion}
In the full-graph expansion all topological distinct graphs with up to $N$ (effective) sites are generated. We define the order of the expansion as the maximum number of sites $N$ taken into account. Note that the sites can be either single spins or a collection of spins like dimers or trimers which function then as effective sites for the graph expansion. In this scheme the number of graphs grows quickly with the order but the Hilbert spaces for ED are typically small for the considered $N$. At the same time it seems impossible to assign typical length scales to graphs which complicates a scaling towards the infinite-order limit. 

\subsection{Rectangular-graph expansion}
In contrast to the full-graph expansion where the number of graphs grows very quickly, the rectangular-graph expansion~\cite{deNeef77,Enting77,Dusuel10,coester15,Kallin13} is restricted to graphs which are rectangular with linear dimensions $L_x$ and $L_y$ so that \mbox{$N=L_x \times  L_y$}. The number of this type of graphs grows much less fast with the number of sites $N$. At the same time the subtraction and embedding procedure becomes trivial for rectangular graphs. The rectangular-graph expansion is by construction especially useful for the square lattice, since rectangular graphs respect the space-symmetries of this lattice. The order is still defined as the number of sites $N$, but now the limiting factor is not the number of graphs, but the ED. In contrast to the full-graph expansion, as shown in Ref.~\onlinecite{Kallin13}, it is possible to introduce the typical length scale $\mathcal{L}_{\rm rect}=\sqrt{N}$ for each order which allows scalings to the infinite-order limit. However, as we detail below, this scaling is still complicated, since the behavior of the rectangular-graph expansion is not smooth with the NLCE order.

\subsection{Arithmetic expansion}
Another scheme, which we call arithmetic expansion, uses again only rectangular graphs, but defines the order as the spatial dimension along the diagonal in a Manhattan-distance sense. The order of an $L_x \times L_y$ rectangular graph is then defined as $L_x+L_y-2$. The $-2$ is chosen, so that the smallest graph, which is a single two-site chain segment, has the order $1$. Therefore the arithmetic order counts the number of links along two of the edges.

For all scalings to the infinite-order limit we introduce the length scale $\mathcal{L}_{\rm arith}$ for the arithmetic expansion. Since the number of sites is not the same for each graph within each order of the arithmetic expansion, we use
\begin{align}
	 \mathcal{L}_{\rm arith} &= \frac{L_x+L_y}{2}
\end{align}
for all scalings, since this corresponds to the average linear extension of the largest quadratic graph in each NLCE order. As we will see below, only even NLCE orders are taken into  account in the scalings and therefore $\mathcal{L}_{\rm arith}$ is an integer number.

\subsection{Square-graph expansion}
Within the rectangular-graph expansion the graphs with the largest spatial extension in one direction are very long chain graphs. This can become problematic for two-dimensional systems, since these one-dimensional graphs might direct the NLCE in the wrong direction. The arithmetic expansion corrects this to a certain extent. Here the maximal chain graph in a given order has the same length as the manhattan distance of the more quadratic graphs. The square-graph expansion goes one step further. Here the order is defined as the maximum number $L_x$ (or $L_y$) of sites into one direction of the rectangular graphs. In order to perform a valid graph expansion all subgraphs of the largest graph must also be taken into account. As a consequence, there are still chain graphs contained in the square-graph expansion, but their length will never be greater than the length of the largest quadratic graph.

Again, as in the arithmetic expansion, the number of sites is not the same for each graph in the expansion. But now it is clear that the quadratic graph with $L \times L$ sites has the maximum number of sites in a given order. Therefore we use $\mathcal{L}_{\rm sq}=L$ as the appropriate length scale for the square-graph expansion when performing scalings to the infinite-order limit.

\section{Edge-fields in NLCE}
\label{Sect:models}
One problem within NLCEs is, that the graphs themself are not ``aware'' of the ordered ground state on the infinite lattice. While the full symmetries of the lattice are restored through the embedding process, single graphs (or subsets of graphs) could show an entirely different behavior compared to the physics of the full lattice, as it has been shown in Ref.~\onlinecite{Ixert15} for the spin-$1$ kagome antiferromagnet. There one-dimensional subsets of graphs behave differently, since the one-dimensional subsystem undergoes a different quantum phase transition in a relevant parameter regime. This situation can be expected generically in quantum many-body systems, especially in the presence of geometric frustration where many different phases compete with each other. As a consequence, it can happen that any partial finite NLCE order displays an erratic behavior and the NLCE essentially breaks down.

To overcome this issue we want to incorporate the following line of reasoning. The NLCE is intended to expand non-perturbatively about the expected quantum phase of a given model which in our case is two-dimensional and long-range ordered. Each graph treated in the NLCE should then contribute the fluctuations specfic to this graph {\it being part} of the ordered phase for the physical quantities of interest, e.g.~the ground-state energy per site. This is only possible if in the calculation on graphs one includes the couplings to the expected ordered state {\it outside} the graph.    

In practice, we assume that the Hamiltonian can be expressed in the following form
\begin{equation}
 \label{Eq::setting}
 \mathcal{H}=\mathcal{H}_0+\lambda\mathcal{V}
\end{equation}
so that $\mathcal{H}_0$ has a symmetry broken ground state \mbox{$\ket{0}\equiv\cdots\ket{0_i}\ket{0_j}\cdots$} which is a product state and is adiabatically connected to the ordered quantum phase expected to be present at finite values of $\lambda$ (often $\lambda=1$ is targeted). Let us mention that the parameter $\lambda$ might be already part of the original model under study or is introduced ``by hand'' corresponding to a deformation of the Hamiltonian to the desired form. 

\begin{figure}
\begin{center}
\includegraphics[width=0.65\columnwidth]{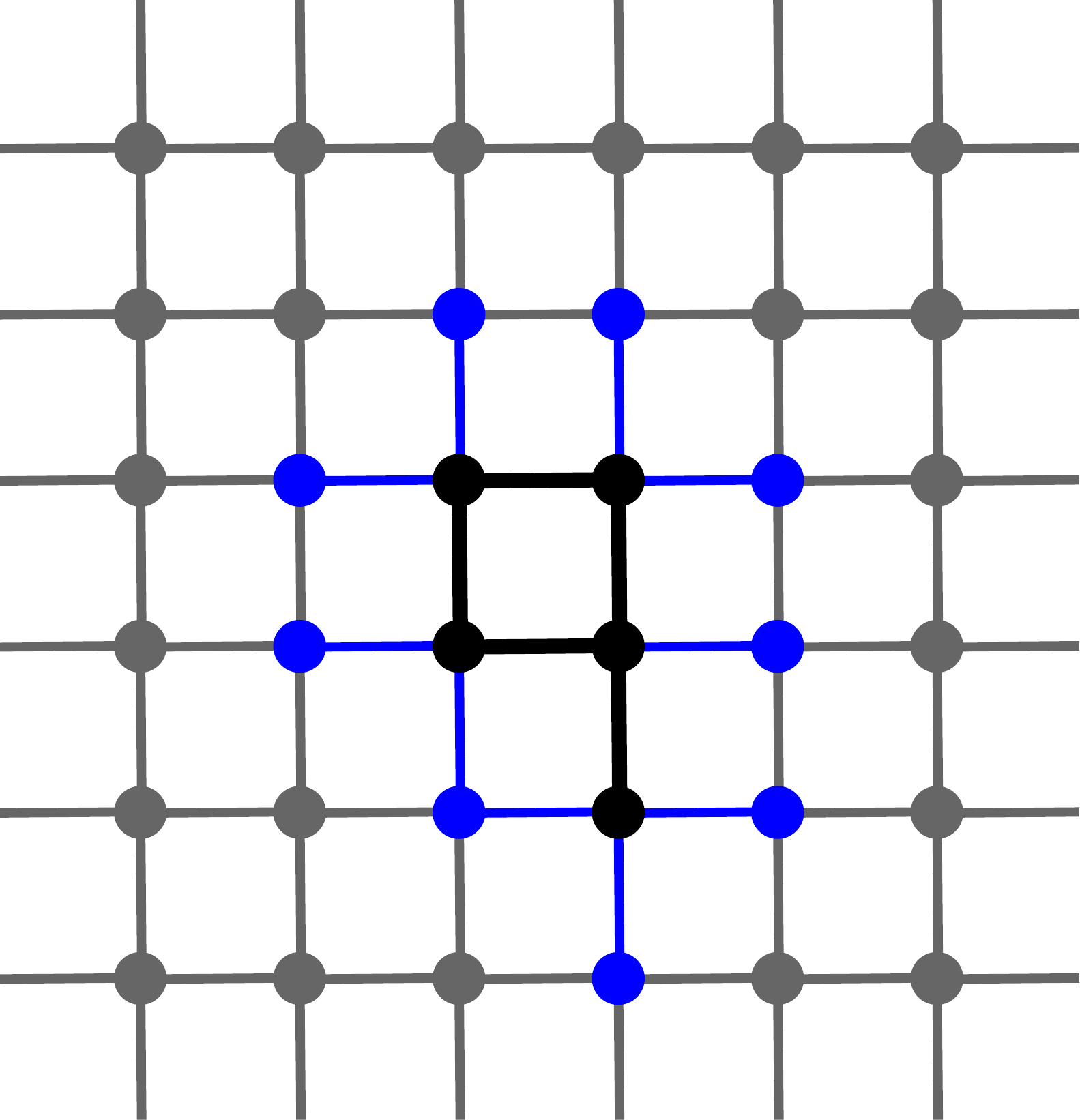}
\end{center}
\caption{(Color online) Illustration of edge-couplings (blue) for one graph (black) on the square lattice (gray).}
\label{fig:EdgeFieldLattice}
\end{figure}

For the NLCE, the elementary site is then chosen according to the product-state structure of $\mathcal{H}_0$. The graphs $\mathcal{G}$ are considered to be surrounded by the state $\ket{0}$. All couplings between a site $i$ of $\mathcal{G}$ and a site $\nu$ outside $\mathcal{G}$ are called the ``edge-couplings'' $\mathcal{O}_\nu\otimes\mathcal{O}_i$ of $\mathcal{G}$. These edge-couplings are illustrated in Fig.~\ref{fig:EdgeFieldLattice} for a particular graph on the square lattice. All couplings within the graph are treated as usual during the ED. However, the operator $\mathcal{O}_\nu\otimes\mathcal{O}_i$ of an edge-coupling reduces to the edge-field 
\begin{equation}
 \bra{0_\nu}\mathcal{O}_\nu\ket{0_\nu}\mathcal{O}_i
\end{equation}
on site $i$ of $\mathcal{G}$. As a consequence, the correct symmetry-breaking is incorporated inside the NLCE scheme. The strength of the total edge-field influencing the results on a graph scales with the perimeter of the graph. As a result, the influence on one-dimensional chain graphs is maximal while it is minimal on the most two-dimensional graphs. This is exactly inline with the above reasoning.  

Let us stress that due to the subtraction and embedding procedure within NLCE the impact of edge-fields becomes less and less for increasing graph sizes, since the size of the bulk of a graph scales much faster with the number of sites $N$ (again, one-dimensional chain graphs are special). In practice, however, the treated graphs are typically not in this limit. We therefore introduce the parameter $K\in\{0,\infty\}$ so that the edge-field  
\begin{equation}
  F_i = -K \mathcal{O}_i
\end{equation}
can be tuned in a flexible fashion. Physically, a value $K\neq \bra{0_\nu}\mathcal{O}_\nu\ket{0_\nu}$ corresponds to a different mean-field product state $\ket{\bar{0}}$. In the limit $K\rightarrow\infty$ the edge-field is so strong that no fluctuations take place on graphs and the system remains in the product state $\ket{0}$ of $\mathcal{H}_0$. In the opposite limit $K\rightarrow 0$, the standard NLCE without edge-fields is recovered with the above mentioned problems. In practice, one expects $K<\bra{0_\nu}\mathcal{O}_\nu\ket{0_\nu}$, since the true ground state of $\mathcal{H}$ contains quantum fluctuations giving smaller values compared to $\bra{0_\nu}\mathcal{O}_\nu\ket{0_\nu}$.

It is important to note, that the contribution of these edge-fields must be subtracted from the ground-state energy within the NLCE. Additionally, as the clusters are getting bigger the contribution of the edge-fields is getting smaller since they are only present at the boundary of the graphs. Therefore the edge-field contribution to the ground-state energy is sub-extensive.

In this work we consider two types of ordered phases which we treat with the edge-field NLCE. The first class are magnetically ordered phases. The quantum ground state can then be considered as a dressed version of the associated classical order, both having the same kind of order parameter corresponding to a finite sublattice magnetization. In this case one can always perform an appropriate sublattice rotation so that the classically ordered state is given by the perfect polarized state where all spins point in $z$-direction. After the sublattice rotation and assuming two-site interactions (a generalization is straightforward), the Hamiltonian can then be written as
\begin{equation}
 \label{Eq::setting_spin_order}
 \mathcal{H}=-J\sum_{i,j} S_i^{z} S_j^{z}+\lambda\mathcal{V}
\end{equation}
which is of the desired form Eq.~\eqref{Eq::setting} with \mbox{$\ket{0}\equiv\ket{\uparrow\cdots\uparrow}$.} The edge-field is therefore a local magnetic field operator on the sites $i$ of the edge of any graph
\begin{align}
	F^{\rm mag}_{i} &= -K S_i^{z} \; . \label{eq:edgeFieldTerm}
\end{align}

The second class of quantum phases are non-magnetic valence bond solids (VBS) which break the translational symmetry of the system. In the most common form the system dimerizes, i.e.~pairs of spins form dimers which themselves order on the lattice. A generalization of this is a trimerization which is important for the spin-$1$ antiferromagnetic Heisenberg model on the kagome lattice.~\cite{Changlani15,Liu15,Li15,Ghosh15,Oitmaa15,Ixert15} Here three spins on a triangle build a low-energy singlet state and it is the trimer entity which is the elementary building block of the ordered state. In both cases (dimerization and trimerization) it is always possible to introduce a parameter $\lambda$ so that for $\lambda=0$ one gets a Hamiltonian $\mathcal{H}_0$ having a ground state of decoupled singlet dimers or trimers. Therefore, the elementary ``sites'' in the NLCE are chosen to be these dimers or trimers.

In contrast to the edge-field NLCE in the spin-ordered phases described above, the Hamiltonian $\mathcal{H}_0$ is completely local in terms of sites. As a consequence, there are no edge-couplings of the form $\mathcal{O}_\nu\otimes\mathcal{O}_i$ for $\lambda=0$. In this case it is mandatory to consider the limit of infinitely small $\lambda$ where edge-couplings appear naturally in second-order perturbation theory in $\lambda$ of the form
\begin{align}
	\left(\ket{s_{\nu}}\bra{s_{\nu}}\right) \otimes \left(\ket{s_{i}}\bra{s_{i}} \right) \quad .
\end{align}  
Here $\ket{s_{i}}\bra{s_{i}}$ is the singlet projector on dimer or trimer $i$ at the edge of a given graph. The edge-fields in the NLCE are then given by
\begin{align}
	F^{\rm VBS}_{i} &=-K \ket{s_{i}}\bra{s_{i}} \quad .
\end{align}

The implementation of edge-fields is now straightforward: we only take ``strong'' graphs~\cite{Oitmaa06} into account. ``Strong'' graphs are graphs where all possible couplings between the sites are set. This is necessary to make the edge-field configuration unique for each graph and not embedding-dependent. For each graph the outgoing couplings of site $i$ then lead to the above defined edge-fields. The next step involves the calculation of extensive ground-state quantities (e.g.~the energy or the sublattice magnetization) via ED. In the last step the considered quantities are reduced and embedded onto the infinite lattice. Remember, that it is important to subtract the edge-field contributions from the ground-state energy. For example, if the reference state is the polarized ferromagnet the subtraction is quite easy, since each edge-coupling leads to a subtraction of $-S \cdot K$, where $S$ is the spin of a single site. This can be seen as a subtraction of single-site contributions, because the edge-fields are local terms which act only on single sites.

The question that remains is, what is the ``optimal'' value of the edge-field strength $K$? In contrast to Ref.~\onlinecite{Kallin13}, in this work we apply the following two different schemes: i) the first is based on a self-consistent solution for the order parameter, e.g.~the sublattice magnetization for the spin-ordered phases. ii) the second targets $K$-values with a minimal sensitivity with respect to the ground-state energy.

i) Physically, it is reasonable to assume that a good choice for $K$ is of the order of the true order parameter of the quantum phase. And indeed, we find that one obtains very good results if $K$ is determined self-consistently as the value of the sublattice magnetization $m_0$ for the spin-ordered phases. We will call this scheme the fixpoint-method (abbreviated by ``fix''), since we are looking for the fixpoint $K=m_0(K)$.

To find this fixpoint we use the secant-method of root finding for each NLCE order, i.e.~for a fixed order we choose two starting values $K_1$ and $K_2$. For both $K$-values we calculate the sublattice magnetization of all graphs which contribute to this NLCE order. Then we subtract and embed them according to the linked-cluster theorem to obtain the values $m_{0}^{(i)}\coloneqq m_0(K_i)$ of the sublattice magnetization. In the final step the secant-method is used to construct an improved choice of $K$. In the \mbox{$(i+1)$th} iteration one has
\begin{align}
	K_{i+1} &=K_i - \frac{K_i-K_{i-1}}{m_{0}^{(i)}-K_i-m_{0}^{(i-1)}+K_{i-1}} \cdot \left( m_{0}^{(i)}-K_i\right)
\end{align}
from which we calculate the corresponding $m_{0}^{(i+1)}$. We stop this iterative process if $\left|K_{i+1}-m_{0}^{(i+1)}\right| < 0.0001$.

ii) The second approach to determine a proper $K$ is to use the criterium of minimal sensitivity with respect to the ground-state energy per site. Physically, this is based on the observation that the effect of $K$ reduces for increasing graph sizes, since it is a subextensive quantity. So in principle the NLCE should converge for each value of $K$ in the limit of large graphs. As a consequence, it makes sense to choose the $K$ in each NLCE order so that the ground-state energy per site depends only minimally on $K$. We therefore check in each NLCE order for a local minimum in the function $e_0(K)$. At this minimum the variation with respect to $K$ is minimal. Note that there exist NLCE orders which do not display a local minimum. In this case case we discard this NLCE order for any further extrapolation to the infinite-order limit. This scheme is called the minimum-method abbreviated by ``min''. 

\section{NLCE scaling}
\label{Sect:scaling}
The NLCE is performed directly in the thermodynamic limit, which is one important aspect compared to other numerical tools. Therefore there are no finite-size effects but each NLCE order corresponds to a different truncation of the real-space fluctuations in the infinite system. Nevertheless, one is interested in a proper scaling of the NLCE data towards the infinite-order limit. Such scalings of NLCEs are a challenging task and a priori no scaling laws are known for NLCEs to the best of our knowledge. In this section we list two different scaling behaviors which we test below: (i) The first is known from conventional finite-size scaling of gapless spin-ordered Heisenberg models \cite{Sandvik97}. (ii) The second stems heuristically from the coupled-cluster method (CCM) \cite{Bishop00} which is, similar to NLCEs, a tool which works directly in the thermodynamic limit.

(i) If we consider a long-range ordered antiferromagnetic ground state in two dimensions with gapless spin-wave excitations, then the finite-size scaling for the ground-state energy per site $e_0$ and the sublattice magnetization $m_0$ is given by \cite{Sandvik97}
\begin{align}
	e_0 (\mathcal{L}) &=e_0 +\frac{a_1}{\mathcal{L}^3}+\frac{a_2}{\mathcal{L}^4}+\ldots\\
        m_0^2 (\mathcal{L}) &=m_0^2 +\frac{b_1}{\mathcal{L}}+\frac{b_2}{\mathcal{L}^2}+\ldots\quad .
\end{align}
Here $\mathcal{L}$ is the linear length of the considered finite system, $a_1,a_2,b_1,b_2$ are fitting parameters, and $\ldots$ denotes higher-order terms in $1/\mathcal{L}$.

(ii) In CCM, the following heuristic scaling laws have been deduced \cite{Bishop00}
\begin{align}
	e_0 (\mathcal{L}) &=e_0 +\frac{\bar{a}_1}{\mathcal{L}^4}+\frac{\bar{a}_2}{\mathcal{L}^8}+\ldots\\
        m_0 (\mathcal{L}) &=m_0 +\frac{\bar{b}_1}{\mathcal{L}^2}+\frac{\bar{b}_2}{\mathcal{L}^4}+\ldots\quad ,
\end{align}
which allows an optimal scaling of CCM data obtained directly in the thermodynamic limit. Here $\bar{a}_1,\bar{a}_2,\bar{b}_1,\bar{b}_2$ are fitting parameters, and $\ldots$ denotes again higher powers in $1/\mathcal{L}$.

In the following we will test these scaling laws of type (i) and (ii) for the edge-field NLCE data squences by identifying the length scale $\mathcal{L}$ with the typical length scales $\mathcal{L}_{\rm rect}$, $\mathcal{L}_{\rm arith}$, and $\mathcal{L}_{\rm sq}$ of the different NLCEs.

\section{Results}
\label{Sect:results}
In the following we present the results obtained by our edge-field NLCE for various models and lattices. We start by benchmarking our approach for the N\'eel-ordered ground state of the spin-1/2 Heisenberg square lattice, which is well-studied and allows us to identify the optimal set ups concerning the type of cluster expansion and the choice of the edge-field $K$. Afterwards, we concentrate on the optimal set ups and discuss the results for the other considered systems. 

\subsection{Spin-$1/2$ square lattice Heisenberg model}\label{Sect:results:squareOneHalf}
The spin-$1/2$ Heisenberg model on the square lattice is given by
\begin{align}
	\mathcal{H} &= J \sum_{\langle i,j\rangle} {\mathbf S}_{i}\cdot{\mathbf S}_{j} \; , \label{eq:heisenbergModel}
\end{align}
where $J>0$ is chosen antiferromagnetic and the sum runs over all pairs of nearest neighbors.
The zero-temperature ground state is the long-range ordered N\'eel state \cite{Reger88,Chakravarty89} which represents a true challenge for any real-space approach due to the infinite correlation length and the gapless spin-wave excitations. On the other side the system is geometrically unfrustrated which is expected to help for the convergence of the NLCE. 

\begin{figure}
\begin{center}
\includegraphics[width=\columnwidth]{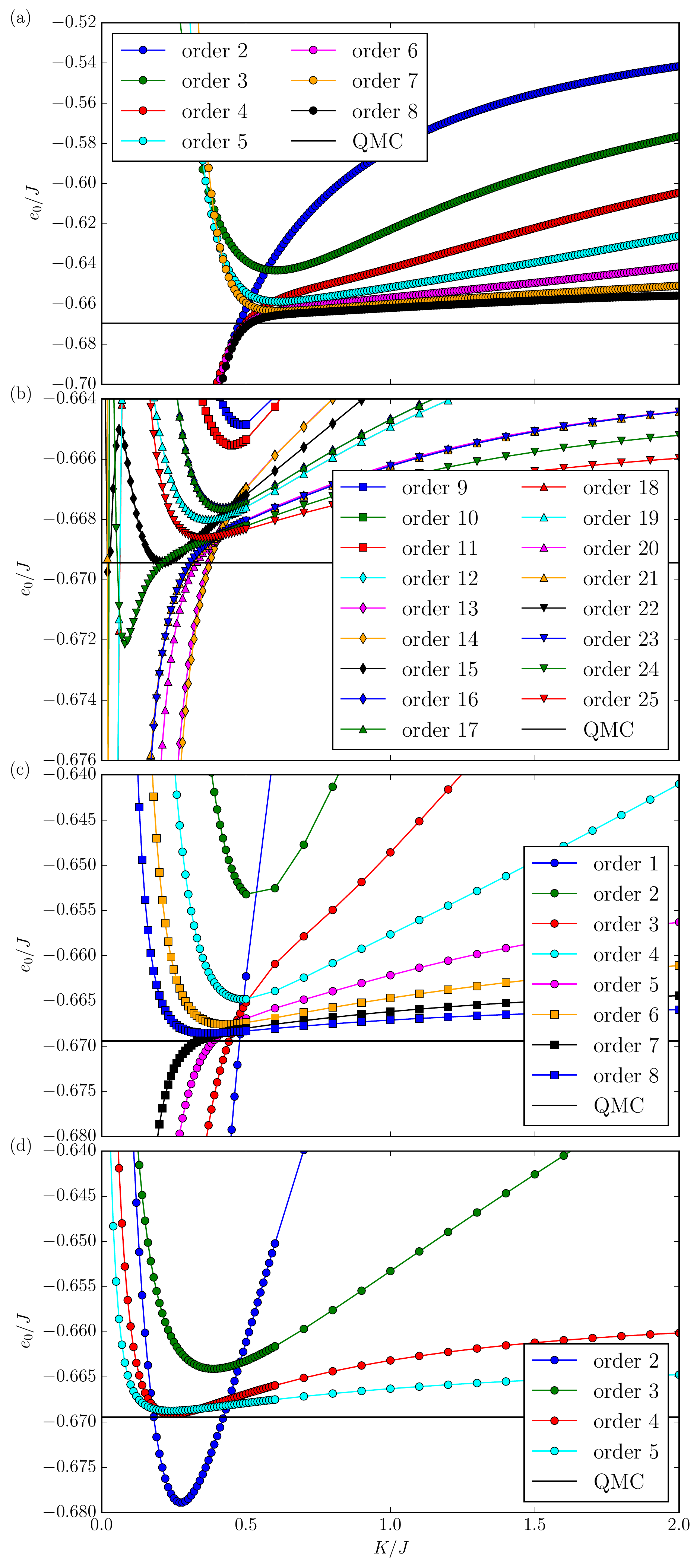}
\end{center}
\caption{(Color online) Ground-state energy per site $e_0/J$ for the spin-$1/2$ square lattice Heisenberg model as a function of the edge-field $K$ using the (a) full-graph, (b) rectangular-graph, (c) arithmetic, (d) square-graph NLCE. Different symbols correspond to different NLCE orders. The lines between symbols are guide to the eyes. The horizontal black line illustrates the QMC-value from Ref.~\onlinecite{Sandvik97}.}
\label{fig:NLCE_squareOneHalf}
\end{figure}

As in standard spin-wave calculations, we perform a sublattice rotation to obtain the following Hamiltonian 
\begin{align}
	\mathcal{H}_{\text{rot}} &= J \sum_{\langle i,j\rangle} \left(- S_i^z S_j^z - S_i^x S_j^x + S_i^y S_j^y \right) \; .
\end{align}
This form is well-suited for our edge-field NLCE. We introduce the parameter $\lambda$ which interpolates between the Ising ($\lambda=0$) and the Heisenberg ($\lambda=1$) limit. This yields the XXZ-Hamiltonian 
\begin{align}
  \label{eq::xxz} 
	\mathcal{H}_{\text{rot}}^{\rm XXZ} &= J \sum_{\langle i,j\rangle} \left[- S_i^z S_j^z+\lambda \left(S_i^y S_j^y - S_i^x S_j^x \right)\right]  \; ,
\end{align}
which is exactly of the form Eq.~\eqref{Eq::setting}. The reference state is then chosen as one of the fully polarized Ising ground states along the $z$-direction and the definition of the edge-fields~\eqref{eq:edgeFieldTerm} are straightforward.

We focus first on the most challenging gapless case $\lambda=1$ for the NLCE. The simpler case of the gapped XXZ-model for $\lambda<1$ is discussed in the next subsection. We start the discussion with the ground-state energy per site $e_0$ and we use the value obtained by quantum Monte Carlo (QMC) simulations~\cite{Sandvik97} $e_0^{\text{QMC}}=-0.669437(5) J$ to gauge our edge-field approach. Note that the QMC value has been obtained by the scaling law of type (i) introduced in Sect.~\ref{Sect:scaling}.

In Fig.~\ref{fig:NLCE_squareOneHalf} (a) we show the ground-state energy per site for the full-graph expansion up to $N=8$ sites as a function of $K/J$. One clearly observes that the NLCE diverges for small $K/J$ (including zero), which already signals the importance to include the effects of the long-range order in the NLCE. In contrary, large values of $K/J$ stabilize the expansion, as expected. If one compares the results to the QMC value, it can be seen that the edge-field NLCE gets closer to the QMC-value with increasing order (which equals to the number of sites $N$ for the full-graph expansion). Still, the expansion is not yet well converged and for $K\lesssim0.6$ an odd-even effect is present. The curves of even order possess no optimal $K$-value, e.g.~no local minima exist. This is different for the odd orders where well-defined minima are present which can serve as the optimal $K$-values.   

\begin{figure}
\begin{center}
\includegraphics[width=\columnwidth]{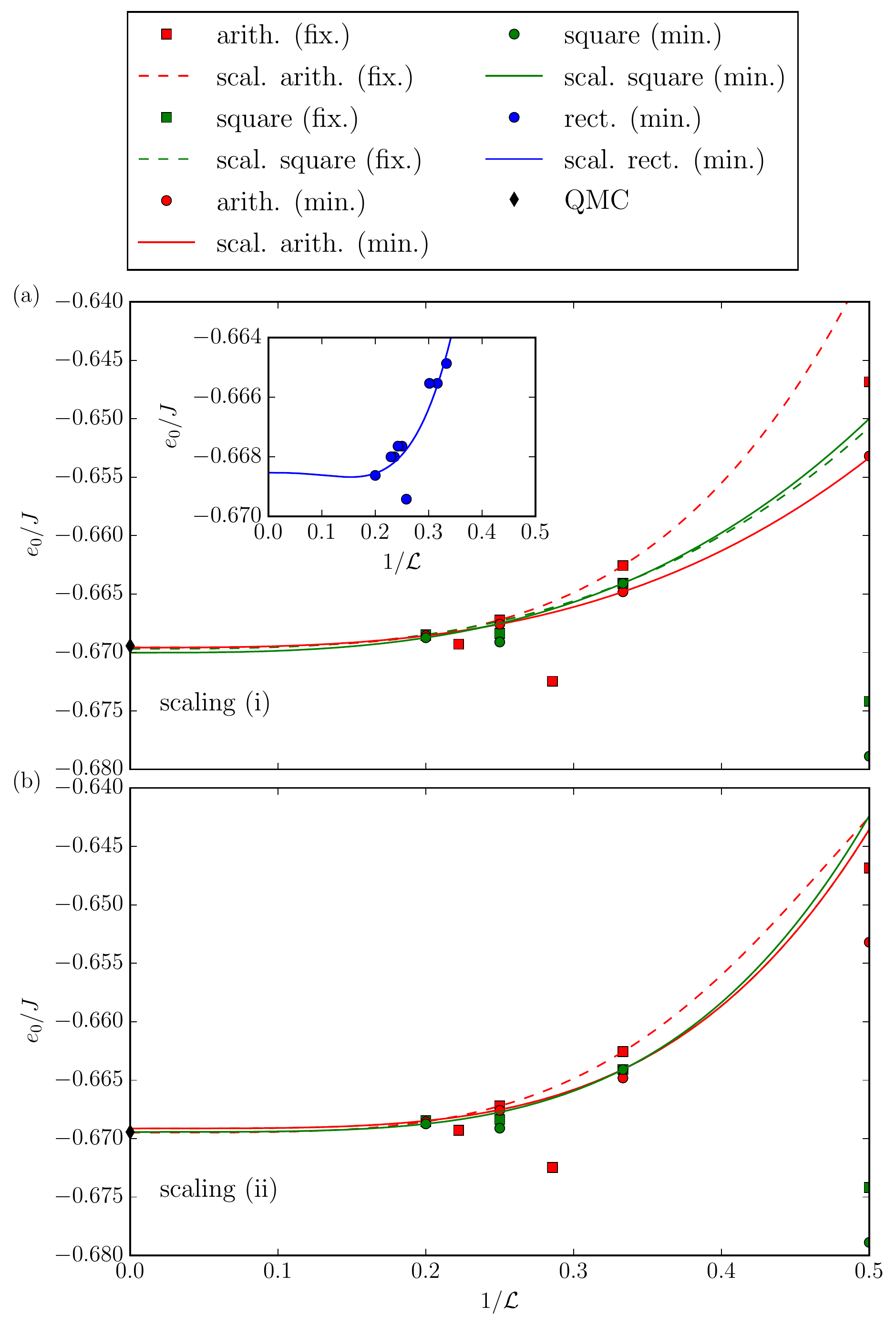}
\end{center}
\caption{(Color online) Upper panel (a) (lower panel (b)) shows scalings of type (i) (of type (ii)) of the ground-state energy per site $e_0/J$ as a function of $1/\mathcal{L}$ for the spin-$1/2$ square lattice as obtained by the minimum and fixpoint method for the arithmetic (red symbols) and square-graph expansion (green symbols). Squares (cirlces) correspond to the values from the fixpoint (minimum) method. The dashed (solid) lines are scalings through data points obtained by the fixpoint (minimum) method. The black diamond depicts the QMC value from Ref.~\onlinecite{Sandvik97}. {\it Inset}: Ground-state energy per site $e_0/J$ as a function of $1/\mathcal{L}$ for the rectangular-graph expansion. Solid line correspond to a scaling of type (i).}
\label{fig:NLCE_squareOneHalf_scaling}
\end{figure}

Next we turn to the results of the rectangular-graph expansion which are displayed in Fig.~\ref{fig:NLCE_squareOneHalf} (b). This expansion is much better converged, which, however, is mainly due to the larger number of sites included in the higher order clusters. Well-defined minima are present as a function of $K$ for odd and even orders alike. If one takes a closer look at these minima in Fig.~\ref{fig:NLCE_squareOneHalf} (b), the minima converge well to the QMC-value. A problematic feature of the rectangular-graph expansion is a proper scaling or extrapolation in the order, which has already been observed in other NLCE studies using rectangular graphs \cite{Kallin13,coester15}. Since $\mathcal{L}_{\rm rect}=\sqrt{N}$ is used to define the typical length scale in the rectangular-graph expansion, energy ``plateaus'' are visible  as can be seen in the inset of Fig.~\ref{fig:NLCE_squareOneHalf_scaling}, i.e.~different orders have almost the same value for the energy and only at certain orders the value changes from one plateau to the one at lower energy. This behavior results due to the fact that certain orders, e.g.~12 and 13, only differ in the contribution from one chain graph (in the example the one with 13 sites) which typically have a very small reduced contribution to the two-dimensional energy per site. As a consequence, the extrapolation to the infinite-order limit is difficult for the rectangular-graph expansion. Nevertheless, a scaling of type (i) (see solid line in the inset of Fig.~\ref{fig:NLCE_squareOneHalf_scaling}) captures the correct magnitude of the energy in the thermodynamic limit.

The arithmetic expansion gets rid of these energy plateaus and leads to very good results as can be seen in Fig.~\ref{fig:NLCE_squareOneHalf} (c). An odd-even effect, as in the full-graph expansion, is again visible and minima as a function of $K$ exist only for the even orders in this expansion. Remarkably, the even orders are the orders in which a quadratic graph is added to the expansion. These graphs are expected to capture the physics of the two-dimensional model in an optimally fashion and are therefore considered to be most important. Consequently, even better results are obtained for the square-graph expansion as shown in Fig.~\ref{fig:NLCE_squareOneHalf} (d). Already the bare minimal values of the ground-state energy per site for order $4$ and order $5$ are very close to the QMC-value. The minimal bare value of order $5$ is $e_0^{\text{min}}=-0.66874 J$, which deviates from the QMC-value by only $0.0007 J$. If one ignores the order $2$ result (which corresponds to the quite small bond and single plaquette clusters) we again expect the presence of an odd-even effect, i.e.~odd and even orders are converging separately to the thermodynamic value. Unfortunately, the order $6$ calculation requires an open cluster with $N=36$ spin-1/2's, which is not possible without the considerable use of symmetries. 

\setlength{\tabcolsep}{10pt}
\begin{table*}[ht]
\begin{tabular}{llllll}
&&\textbf{Spin-$1/2$ square lattice}&&&\\[4pt]
Method 		& Extrapolation 					& $e_0/J$ 		& $|e_0-e_0^\text{QMC}|/J$	& $m_0$ 			& $m_0-m_0^\text{QMC}$ \\ \hline\\[-6pt]
arith. (fix.)		&	scaling (i) ($4$, $6$, $8$)			& $-0.669567$		& $0.00013$				& $0.3077$		& $0.0007$ \\[4pt]
arith. (fix.)		&	scaling (ii) ($4$, $6$, $8$)			& $-0.669486$		& $0.000049$				& $0.3220$		& $0.015$ \\[4pt]
arith. (fix.)		&	Wynn ($4$, $6$, $8$)			& $-0.669081$		& $0.00036$				& $0.3319$		& $0.025$ \\[4pt]
square (fix.)	&	scaling (i) ($3$, $5$)				& $-0.669692$		& $0.00025$				& $0.3038$		& $0.0032$ \\[4pt]
square (fix.)	&	scaling (ii) ($3$, $5$)				& $-0.669134$		& $0.00030$				& $0.3336$		& $0.027$ \\[4pt]
square (fix.)	&	Wynn ($3$-$5$)				& $-0.668482$		& $0.00096$				& $0.3476$		& $0.041$ \\[4pt]
arith. (min.)	&	scaling (i) ($4$, $6$, $8$)			& $-0.669588$		& $0.00015$				& $0.3059$		& $0.0011$ \\[4pt]
arith. (min.)	&	scaling (ii) ($4$, $6$, $8$)			& $-0.669301$		& $0.00014$				& $0.3316$		& $0.025$ \\[4pt]
arith. (min.)	&	Wynn ($4$, $6$, $8$)			& $-0.669102$		& $0.00034$				& $0.3277$		& $0.021$ \\[4pt]
square (min.)	&	scaling (i) ($3$, $5$)				& $-0.670022$		& $0.00058$				& $0.2877$		& $0.019$ \\[4pt]
square (min.)	&	scaling (ii) ($3$, $5$)				& $-0.669433$		& $0.000004$				& $0.3252$		& $0.018$ \\[4pt]
square (min.)	&	Wynn ($3$-$5$)				& $-0.668765$		& $0.00067$				& $0.3435$		& $0.037$ \\[4pt]\hline\hline\\

&&\textbf{Spin-$1$ square lattice}&&&\\[4pt]
Method 		& Extrapolation 					& $e_0/J$ 		& $|e_0-e_0^\text{CCM}|/J$	& $m_0$ 			& $m_0-m_0^\text{CCM}$ \\ \hline\\[-6pt]
square (fix.)	&	scaling (i) ($3$, $4$)				& $-2.32779$		& $0.0020$				& $0.8112$		& $0.017$ \\[4pt]
square (fix.)	&	scaling (ii) ($3$, $4$)				& $-2.32645$		& $0.0033$				& $0.8387$		& $0.045$ \\[4pt]
square (min.)	&	scaling (i) ($3$, $4$)				& $-2.33133$		& $0.0016$				& $0.7630$		& $0.031$ \\[4pt]
square (min.)	&	scaling (ii) ($3$, $4$)				& $-2.32992$		& $0.00017$				& $0.8057$		& $0.012$ \\[4pt]\hline\hline
\end{tabular}
\caption{Comparison of ground-state energies per site $e_0/J$ and sublattice magnetizations $m_0$ on the spin-$1/2$ and spin-$1$ square lattice. The determination method for $K$ is denoted in brackets after the used graph-expansion method (min. for the minimum method and fix. for the fixpoint method). The used orders for the extrapolation are stated after the extrapolation method.}
\label{tab:squareOneHalfEnergy}
\end{table*}

\begin{figure}
\begin{center}
\includegraphics[width=\columnwidth]{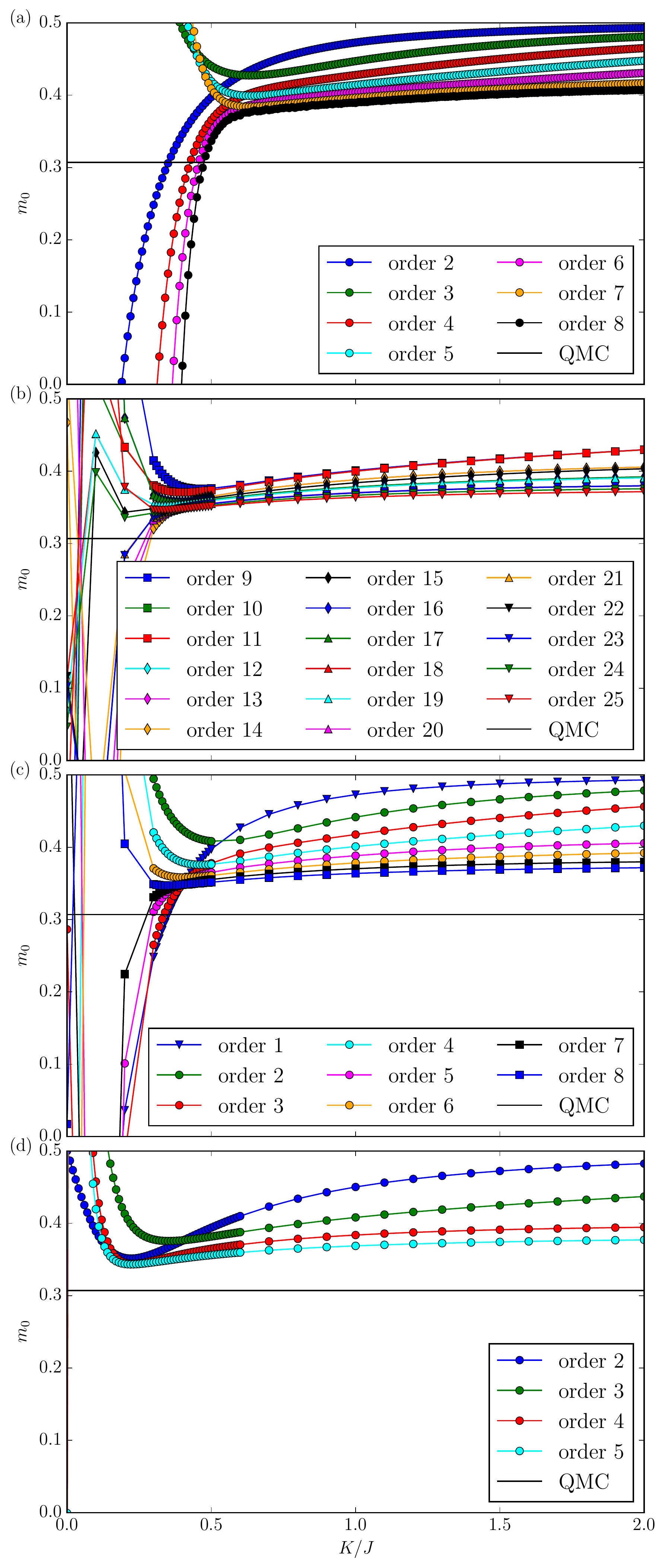}
\end{center}
\caption{(Color online) Sublattice magnetization $m_0$ for the spin-$1/2$ square lattice Heisenberg model as a function of the edge-field $K$ using the (a) full-graph, (b) rectangular-graph, (c) arithmetic, (d) square-graph NLCE. Different symbols correspond to different NLCE orders. The lines between symbols are guide to the eyes. The horizontal black line illustrates the QMC-value from Ref.~\onlinecite{Sandvik97}.}
\label{fig:NLCE_squareOneHalf_mag}
\end{figure}

Next we take the $K$-values of the well-defined minima as well as the fixpoint $K$-values for each NLCE order to perform scalings of type (i) and (ii) as well as to apply the Wynn-algorithm (see e.g. Ref.~\onlinecite{Tang13} and~\onlinecite{Guttmann89}) in order to obtain even better estimates of the ground-state energy per site in the infinite-order limit. Note that another option to extrapolate NLCE data sequences has been recently formulated in Ref.~\onlinecite{Coester2016}. However, this scheme is most powerful to extract critical properties which we do not consider in this work.  The obtained values for the arithmetic and square-graph NLCE for the different extrapolations are listed in Tab.~\ref{tab:squareOneHalfEnergy} and are illustrated in Fig.~\ref{fig:NLCE_squareOneHalf_scaling}. In accordance with the explanations given above, the quality of the full- and the rectangular-graph expansion is not as high as the other two NLCEs. In the full-graph expansion the maximal order is clearly not competitive and it would be also interesting to push this expansion to higher orders. In contrast, in the rectangular-graph expansion the plateau-effect complicates a proper extrapolation in the order. We therefore do not show these results. We want to stress that not every order has a well-defined fixpoint, since sometimes such points are absent as can be seen in Fig.~\ref{fig:NLCE_squareOneHalf_mag}. Note that the orders which have a well-defined minimum of the magnetization also have a fixpoint. 

The scalings of type (i) and (ii) are explicitly shown in Fig.~\ref{fig:NLCE_squareOneHalf_scaling}. Here we scaled through the ``best'' NLCE orders. ``Best'' implies that the involved orders should be as large as possible by respecting odd-even effects. Here we took the three largest even orders $4$, $6$, and $8$ in the arithemtic expansion while we took orders $3$ and $5$ for the square-graph expansion. The used orders are also given in Tab.~\ref{tab:squareOneHalfEnergy}. 

\begin{figure}
\begin{center}
\includegraphics[width=\columnwidth]{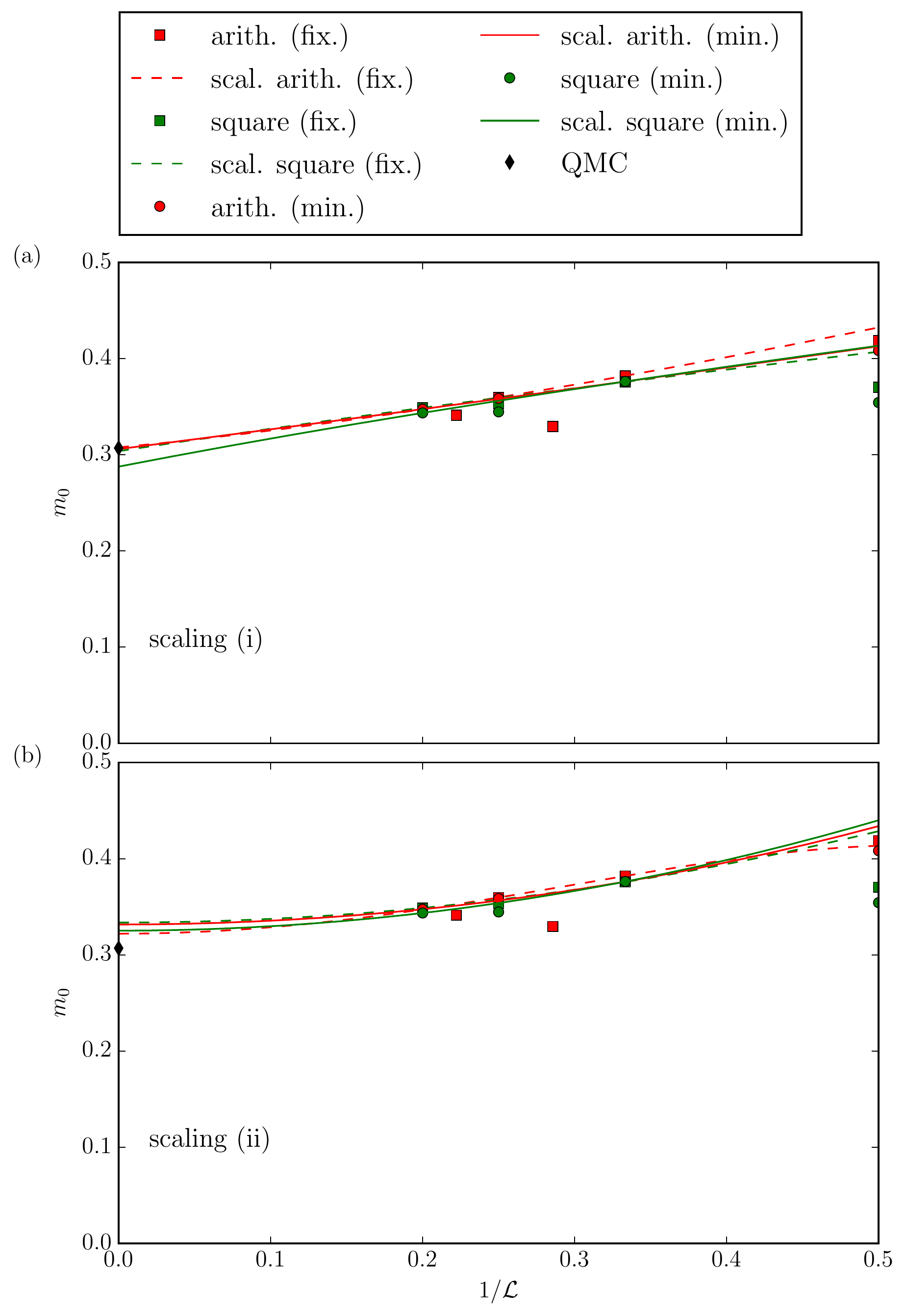}
\end{center}
\caption{(Color online) Upper panel (a) (lower panel (b)) shows scalings of type (i) (of type (ii)) of the sublattice magnetization $m_0$ as a function of $1/\mathcal{L}$ for the spin-$1/2$ square lattice as obtained by the minimum and fixpoint method for the arithmetic (red symbols) and square-graph expansion (green symbols). Squares (circles) correspond to the values from the fixpoint (minimum) method. The dashed (solid) lines are scalings through data points obtained by the fixpoint (minimum) method. The black diamond depicts the QMC value from Ref.~\onlinecite{Sandvik97}.}
\label{fig:NLCE_squareOneHalf_mag_scaling}
\end{figure}

Globally, all schemes give values close to the QMC value improving the best bare NLCE values by roughly one order of magnitude. We observe that the fixpoint scheme as well as the local $K$-minima approach give similarly good results for both NLCEs. At the same time both type of scalings perform slightly better than the values from Wynn extrapolation. However, the quality of both scaling types is almost the same so that a comparison is complicated. 

Next we discuss the behavior of the ground-state sublattice magnetization $m_0$ which corresponds to the order parameter of the N\'eel state. We stress that this is only possible if one performs the NLCE inside the symmetry-broken phase which is realized in our approach due to the presence of the edge-fields. As a reference, we again take the QMC value from Ref.~\onlinecite{Sandvik97} which is $m_0^{\text{QMC}}=0.3070(3)$. The obtained edge-field NLCE results for $m_0$ as a function of $K$ are displayed in Fig.~\ref{fig:NLCE_squareOneHalf_mag} for the four different cluster expansions. 

All general features discussed for the ground-state energy per site are also present for the sublattice magnetization. Most importantly, well-defined minima exist at roughly the same $K$-values as for the ground-state energy per site. The convergence of the bare NLCE values to the QMC-value is not as good as for the ground-state energy, but the general trend is the same. We observe that the bare values at the local minima and the fixpoints are typically larger than the QMC-value. This originates from the fact that the fully classical reference state likely yields a too large edge-field for the considered clusters.  

Along the same lines as for the ground-state energy per site, we performed scalings of type (i) and (ii) as shown in Fig.~\ref{fig:NLCE_squareOneHalf_mag_scaling}. Note that for the scaling we took the $K$-values where energy is minimal and not the local minima of the magnetization in order to get a consistent scheme. As before, for the fixpoint-method the scaled values are of course taken from the sublattice magnetization. The obtained scaled values for $m_0$ are also listed in Tab.~\ref{tab:squareOneHalfEnergy} together with their deviation from the QMC-value. These differences are larger compared to the ones for the ground-state energy per site, which is however expected, since the sublattice magnetization is considerably more sensitive. Typically, we obtain a satisfactory agreement with QMC having a difference of the order $0.01 J$. We find that the arithmetic expansions gives better results compared to the square-graph expansion. However, this is likely due to the fact that only two data points are included in the square-graph expansion which enhances the uncertainty of the scaling procedure. If one compares the two types of scaling, then it is apparent that type (i) yields more convincing results than type (ii), which typically overshoots the QMC value.  

Altogether, the edge-field NLCE gives quantitative results for the ground-state energy per site and the sublattice magnetization for the ordered N\'eel state of the square lattice spin-1/2 Heisenberg model having an infinite correlation length. We have seen that the fixpoint and the local $K$-minima approaches give both comparable and convincing results. In the following we will focus on the scaling behavior in all other applications of the edge-field NLCE below. We therefore do not apply the Wynn algorithm anymore, but restrict the discussion to the two types of scaling laws. 

\subsection{Spin-$1/2$ XXZ-model on the square lattice}

In this subsection we discuss the edge-field NLCE for $\lambda\in [0,1 ]$ in Eq.~\eqref{eq::xxz} corresponding to the XXZ model on the square lattice. This model interpolates between the Ising limit $\lambda=0$ and the just discussed Heisenberg model for $\lambda=1$. We stress that for $\lambda<1$ the system is gapped and the correlation length is finite. As a consequence, one expects the edge-field approach to converge faster with the order compared to the gapless Heisenberg model and to agree with perturbative SE for small $\lambda$. This is exactly what we find.

\begin{figure}
\begin{center}
\includegraphics[width=\columnwidth]{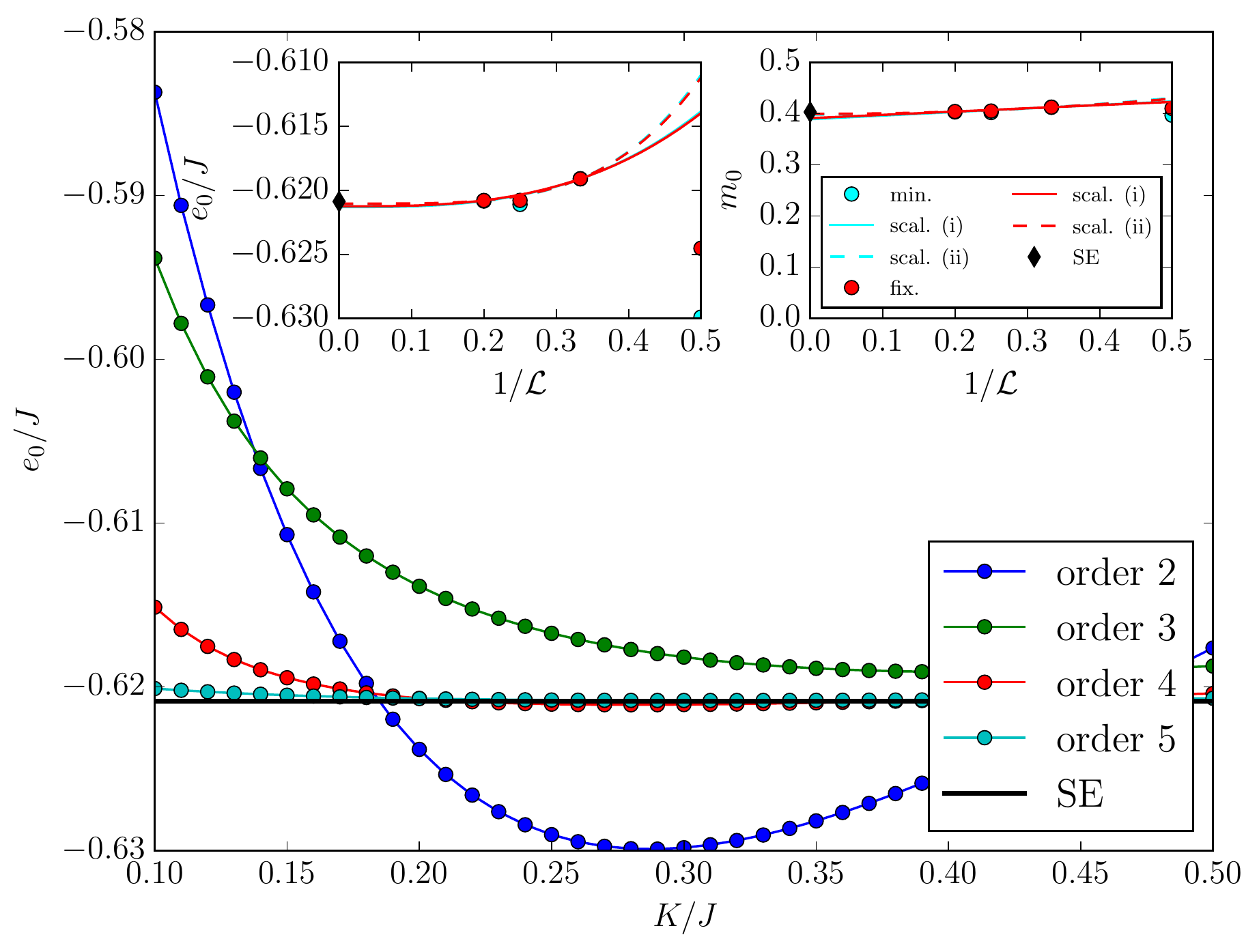}
\end{center}
\caption{(Color online) Ground-state energy per site $e_0/J$ for the spin-$1/2$ XXZ-model on the square lattice for $\lambda=0.85$ using the square-graph NLCE. Different colors correspond to different orders of graph expansion. The horizontal black line illustrates the bare value obtained by SE~\cite{Singh89, Weihong91, Dusuel10} up to order $14$.
\textit{Left inset:} Scaling of type (i) (red solid line) and type (ii) (red dashed line) of the $e_0$-values obtained by the minimum- (cyan circles) or fixpoint-method (red circles) as a function of $1/\mathcal{L}$. \textit{Right inset:} Scalings of type (i) (solid lines) and type (ii) (dashed lines) of the corresponding $m_0$-values as a function of $1/\mathcal{L}$. Cyan (red) symbols/lines correspond to the values from the minimum-method (fixpoint-method), whereas Diamonds refer to the bare SE result.}
\label{fig:NLCE_squareOneHalf_XXZ_lambda0.85}
\end{figure}
\begin{figure}
\begin{center}
\includegraphics[width=\columnwidth]{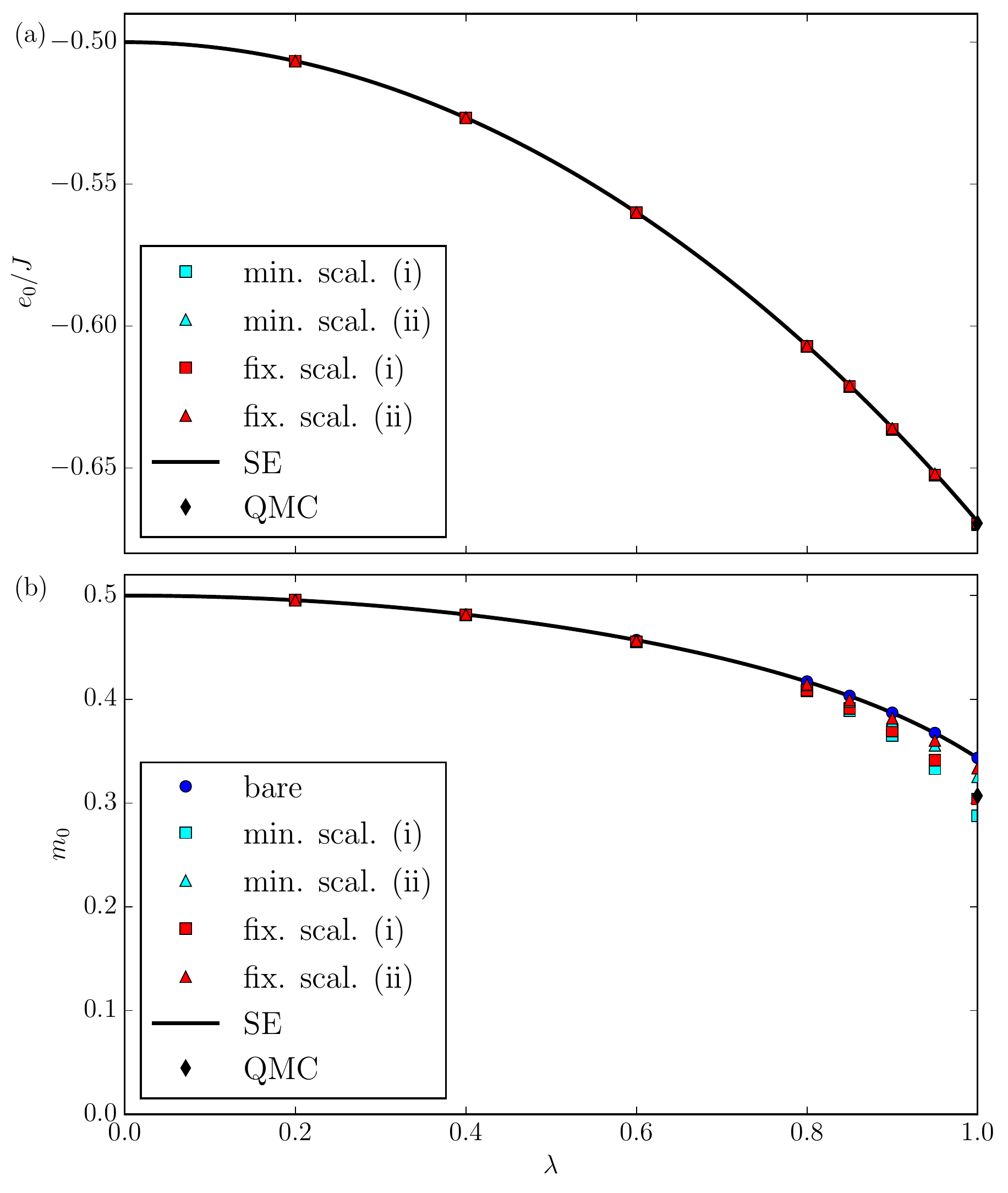}
\end{center}
\caption{(Color online) (a) Ground-state energy per site $e_0/J$ and (b) sublattice magnetization $m_0$ for the XXZ-model on the square lattice obtained by scalings of type (i) (squares) and type (ii) (triangles) of the square-graph expansion. Cyan symbols result from the minimum-method and red symbols from the fixpoint-method. Circles correspond to the bare order $5$ NLCE, the solid line is the bare high-order SE in $\lambda$ up to order $14$ from Refs.~\onlinecite{Singh89, Weihong91, Dusuel10}, and the diamond refers to the QMC-value~\cite{Sandvik97} for the Heisenberg point $\lambda=1$.}
\label{fig:NLCE_squareOneHalf_XXZ}
\end{figure}

In this part we focus on the square-graph expansion, since this expansion performed best for the Heisenberg case. A representative plot of the ground-state energy per site and the sublattice magnetization for the specific value $\lambda=0.85$ is shown in Fig.~\ref{fig:NLCE_squareOneHalf_XXZ_lambda0.85}. One clearly observes that already the bare NLCE values are well converged. Again, as for the Heisenberg model, scalings of type (i) and type (ii) are performed through the orders $3$ and $5$ minimal values of the ground-state energy per site which further improves the quality of the NLCE data (see insets in Fig.~\ref{fig:NLCE_squareOneHalf_XXZ_lambda0.85}).  

Scaled NLCE results for $e_0$ and $m_0$ on the full $\lambda$-axis are presented in Fig.~\ref{fig:NLCE_squareOneHalf_XXZ}. These results are compared to high-order SE in $\lambda$ from Refs.~\onlinecite{Singh89, Weihong91, Dusuel10}. Note that the highest NLCE order 5 of the square-graph expansion contains the order eight perturbation theory in $\lambda$ exactly (this is actually true for all four different NLCEs). It is therefore no surprise but in fact mandatory that SE and NLCE are in quantitative agreement for small $\lambda$. Genercally, for larger values of $\lambda$ this must not be the case, since the bare SE need not converge while the NLCE could still be convergent. However, for the unfrustrated XXZ model on the square lattice, all SE are monotonous and therefore even the bare perturbative series yields satisfactory results. This will be different for the same model on the triangular lattice, where one finds alternating series due to the geometric frustration.  

Overall, the obtained NLCE results using edge-fields give convergent and satisfactory results on the full $\lambda$-axis for the unfrustrated XXZ model on the square lattice. Next we investigate the same model for larger spins one. 

\subsection{Spin-$1$ square lattice Heisenberg model}

The edge-field NLCE of the spin-1/2 XXZ-model on the square lattice is a systematic and non-perturbative expansion about the long-range ordered classical N\'eel state. All quantum fluctuations contained on the clusters in a given NLCE order are taken fully into account. As discussed above, the most challenging case for the NLCE is the gapless Heisenberg model for $\lambda=1$ due to the diverging correlation length. In this subsection we focus again on the Heisenberg model but enlarge the spin value to one. The system is therefore still long-range ordered and gapless. The N\'eel-ordered reference state is unchanged and one expects that the edge-field NLCE converges faster, since larger spins display smaller quantum fluctuations and the edge-fields themselves originate from a mean-field decoupling assuming the classical N\'eel-ordered reference state outside the clusters under investigation. At the same time the ED on clusters is harder, since the local Hilbert space of spins one is larger. As a consequence, we do not reach the same cluster sizes as for spins 1/2.


\begin{figure}
\begin{center}
\includegraphics[width=\columnwidth]{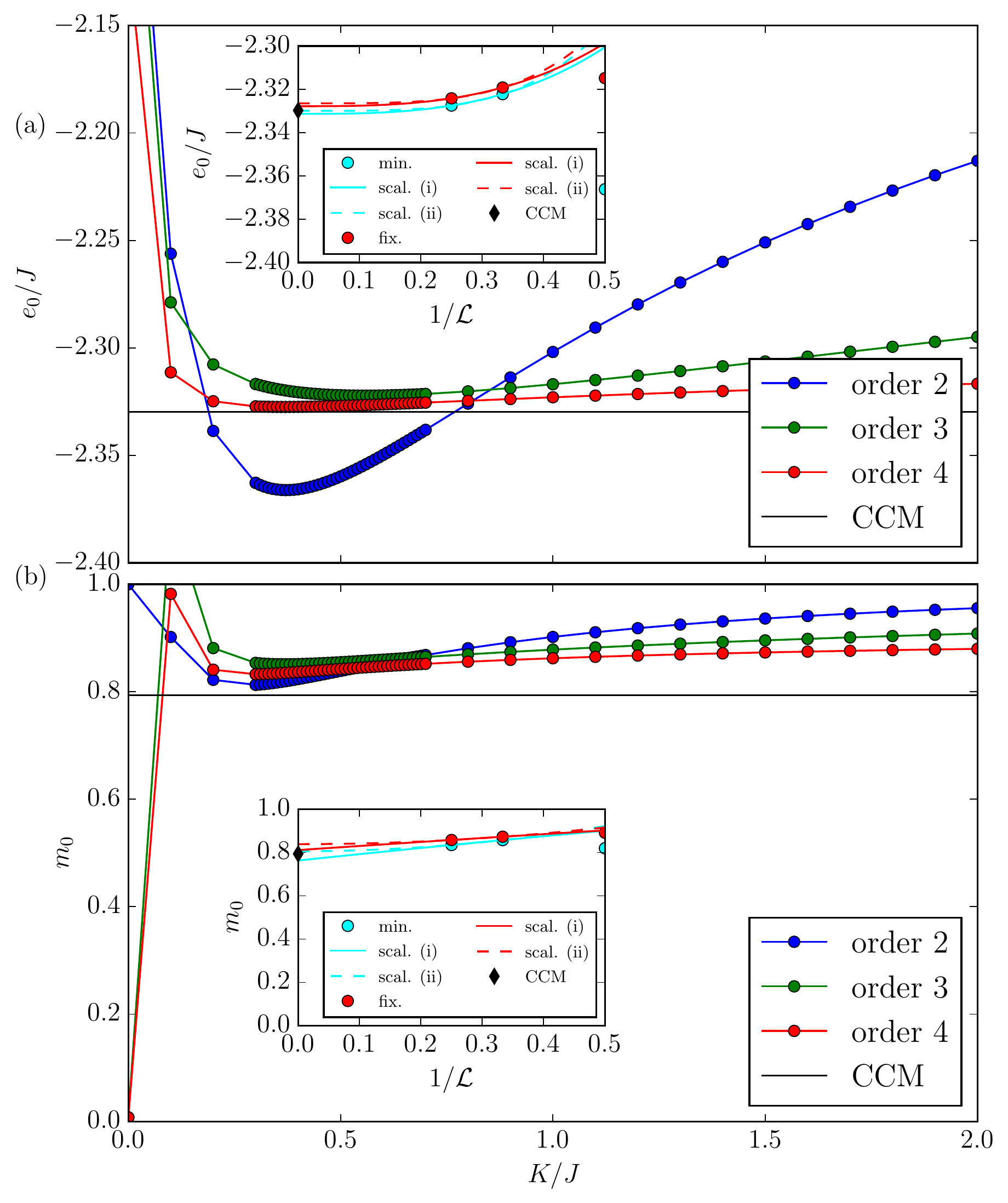}
\end{center}
\caption{(Color online) (a) Ground-state energy per site $e_0/J$ and (b) sublattice magnetization $m_0$ for the spin-$1$ Heisenberg model on the square lattice obtained with edge-field NLCE using the square-graph expansion. Different colored symbols correspond to different orders of the graph expansion. The lines between them are guide to the eyes. The insets show scalings of type (i) (solid lines) and type (ii) (dashed lines) through the NLCE orders $3$ and $4$ (cyan circles for the minimum- and red circles for the fixpoint-method) of the ground-state energy per site or the sublattice magnetization as a function of $1/\mathcal{L}$. The black diamond depicts the CCM value from Ref.~\onlinecite{Li11}.}
\label{fig:NLCE_squareOne_squareExpansion}
\end{figure}

In the following we focus on the square NLCE, which was the best NLCE in the spin-$1/2$ case. Due to the larger spin-one Hilbert space, we reach order four in the square NLCE, i.e.~the maximal cluster contains 16 sites. The corresponding NLCE results for the ground-state energy per site $e_0/J$ and the sublattice magnetization $m_0$ are displayed in Fig.~\ref{fig:NLCE_squareOne_squareExpansion}. A scaling of the bare NLCE data has also been performed as for the spin-1/2 case and yields again convincing results which are listed in the lower panel of Tab.~\ref{tab:squareOneHalfEnergy}. As a reference, we compare our NLCE results to the values $e_0^{\text{CCM}}=-2.32975 J$ and $m_0^{\text{CCM}}=0.7938$ obtained with the CCM in Ref.~\onlinecite{Li11}. The reduced quantum fluctuations in the spin-one case can be readily seen from the rather large value of the sublattice magnetization. 

One observes that already the bare NLCE orders converge very well. The only exception, as for the spin-1/2 case, is the lowest order two which does not fit in the general trend of the NLCE. As a consequence, we discard this order from any scaling of the NLCE data. Already the  bare NLCE order four yields the value $e_0^{\text{min}}=-2.32744 J$, which differs only by $0.00231 J$ from the CCM-value~\cite{Li11}. This means that the order four NLCE with up to $16$ sites for spin one is as good as the order five NLCE for spins-1/2 with up to $25$ sites. For the spin-$1$ case, the quality of the NLCE is only slightly improved due to scaling (see Tab.~\ref{tab:squareOneHalfEnergy}), which is mainly due to the lower order of the NLCE.

As for the spin-1/2 case, the quality of the sublattice magnetization is slightly lower as can be seen from Fig.~\ref{fig:NLCE_squareOne_squareExpansion}~(b) as well as from Tab.~\ref{tab:squareOneHalfEnergy} where we list the scaled values of the different NLCEs. The scaling of the sublattice magnetization is again performed as described for the spin-$1/2$ case (see inset of Fig.~\ref{fig:NLCE_squareOne_squareExpansion}~(b)). Again, as for the ground-state energy per site, the bare NLCE values converge smoothly to the CCM-value. We observe that the scaling of type (i) results in a too low value compared to the CCM-value \cite{Li11} while the value deduced from scaling of type (ii) is too large. These discrepancies originate from the uncertainty of the scaling, since we only have the two points from order three and four, and we expect also an even-odd effect in the NLCE as for the spin-1/2 NLCE data.  

Altogether, the edge-field NLCE performs also well for the spin-$1$ Heisenberg model despite the fact that the maximal NLCE order is reduced compared to the spin-1/2 case due to the larger Hilbert space. 

\subsection{Spin-$1/2$ triangular lattice Heisenberg model}

The Heisenberg model on the square lattice is unfrustrated for any value of the spin. This is expected to help the NLCE to converge with increasing NLCE order, since the bare reference state of $\mathcal{H}_0$ is unfrustrated on each cluster and therefore no competing ground states are close in energy for the vast majority of clusters. In contrast, on a geometrically frustrated system like the antiferromagnetic Heisenberg model on the triangular lattice the situation is more complex and it is a priori not clear how the edge-field NLCE performs.  

\begin{figure}
\begin{center}
\includegraphics[width=\columnwidth]{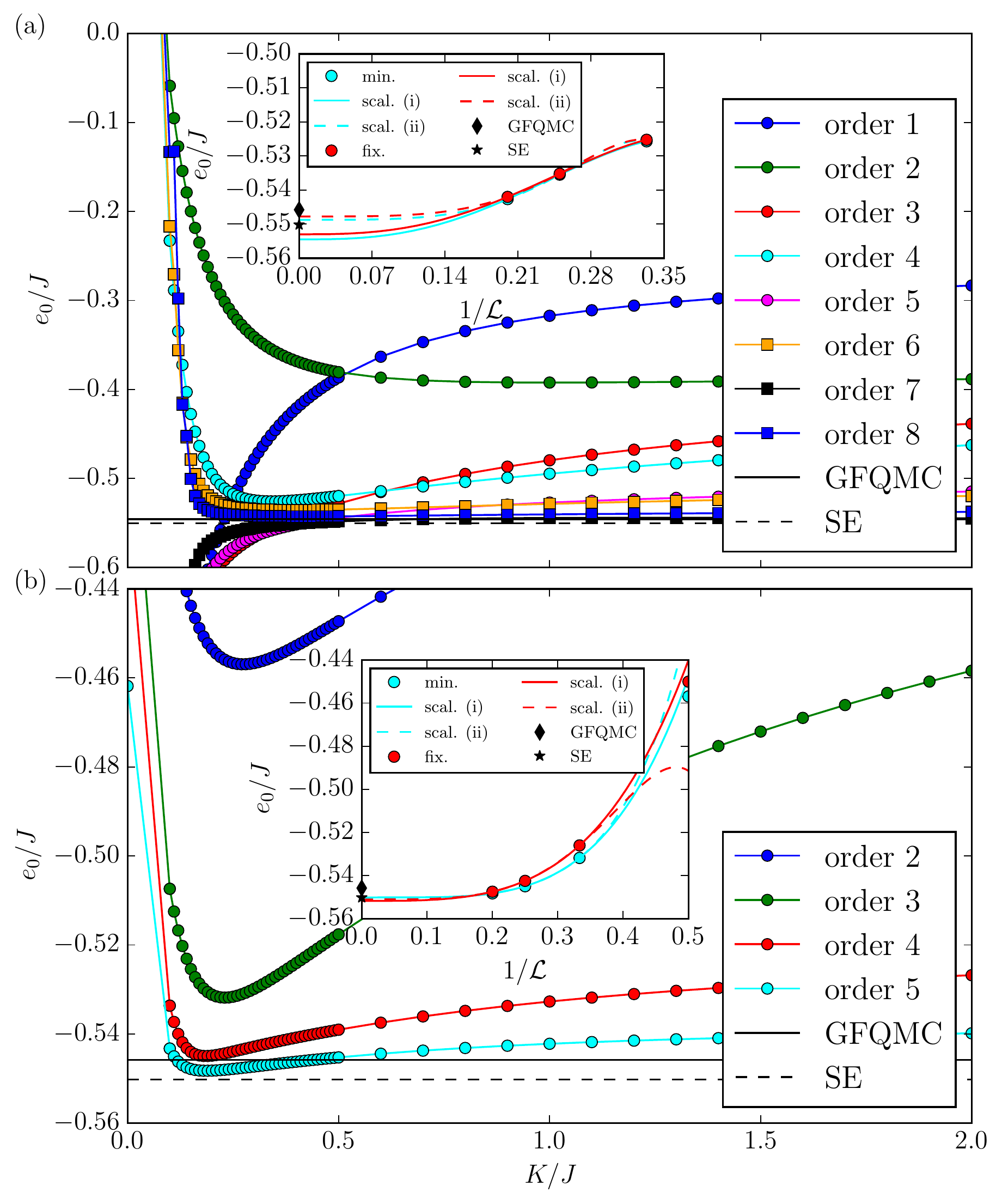}
\end{center}
\caption{(Color online) Ground-state energy per site $e_0/J$ for the (a) arithmetic expansion and (b) square-graph expansion for the spin-$1/2$ triangular lattice. Different colored symbols correspond to differend orders of the graph expansion. The lines between them are guide to the eyes. Insets show scalings of type (i) (solid lines) and type (ii) (dashed lines) as a function of $1/\mathcal{L}$ through the minimal (cyan circles) and fixpoint (red circles) NLCE values. The black diamond depicts the GFQMC value~\cite{Capriotti99} and the star corresponds to the SE value~\cite{Zheng06}.}
\label{fig:NLCE_triangularOneHalf_arithmeticSquare}
\end{figure}

As for the Heisenberg model on the square lattice, the ground state on the triangular lattice breaks the continuous SU(2) symmetry and the system possesses gapless Goldstone modes. The specific order is given by the so-called three-sublattice $120^\circ$ order for any value of the spin. It is therefore again possible to view the quantum ground state of the triangular Heisenberg model as a dressed version of the classical $120^\circ$ order of the corresponding Ising model on the same lattice. As a consequence, the edge-field NLCE for the $120^\circ$ quantum order can be applied along the same lines as for the N\'eel order on the square lattice.  

\begin{figure}
\begin{center}
\includegraphics[width=\columnwidth]{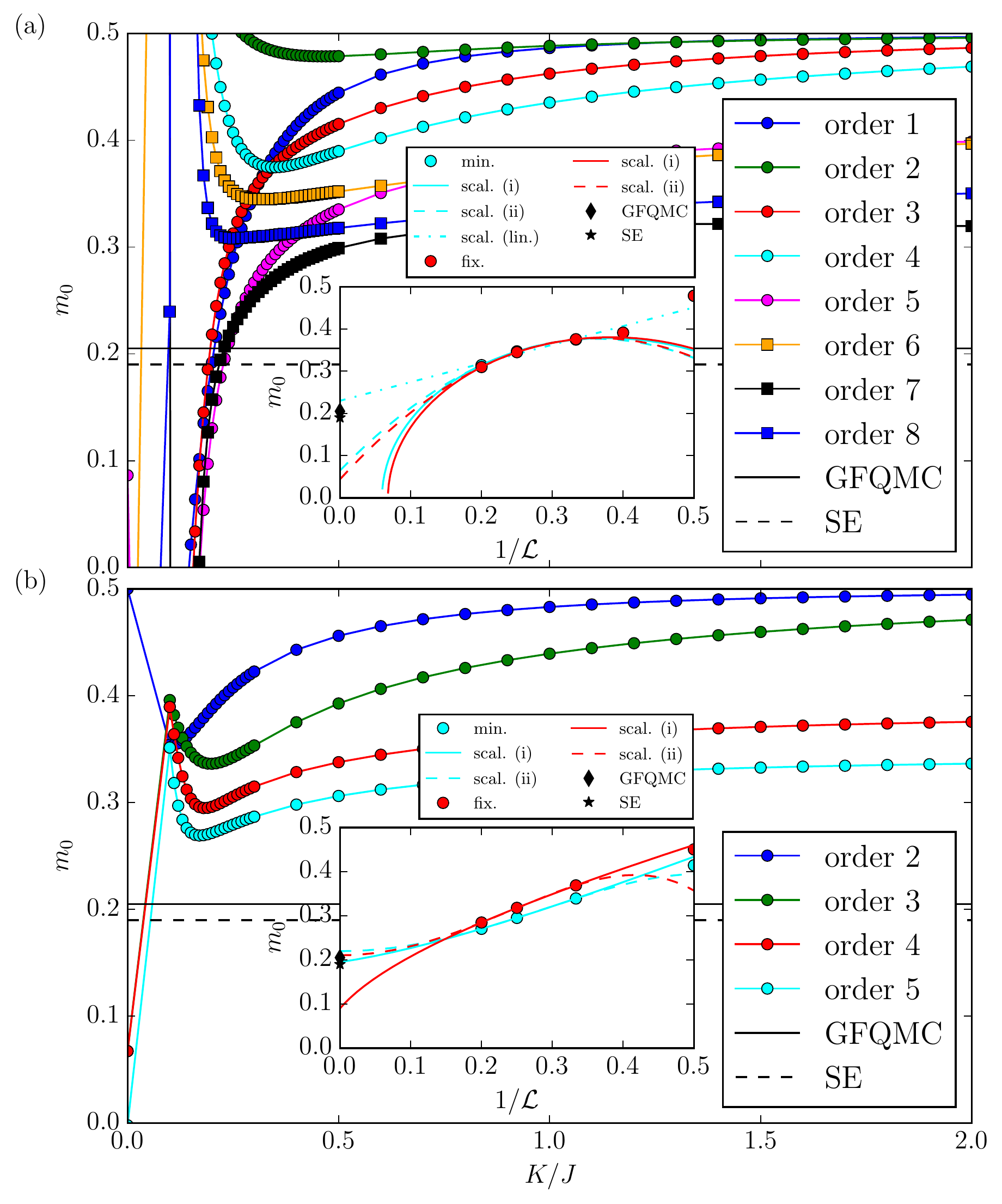}
\end{center}
\caption{(Color online) Sublattice magnetization $m_0$ for the (a) arithmetic expansion and (b) square-graph expansion for the spin-$1/2$ triangular lattice. Different colored symbols correspond to differend orders of the graph expansion. The lines between them are guide to the eyes. Insets show scalings of type (i) (solid lines) and type (ii) (dashed lines) as a function of $1/\mathcal{L}$ through the minimal (cyan circles) and fixpoint (red circles) NLCE values. For the arithmetic expansion we additionally display a linear fit as a dashed-dotted line in the upper inset. The black diamond depicts the GFQMC value~\cite{Capriotti99} and the star corresponds to the SE value~\cite{Zheng06}.}
\label{fig:NLCE_triangularOneHalf_arithmeticSquare_mag}
\end{figure}

If one performs the appropriate sublattice rotation for the $120^\circ$ ordered state to the Heisenberg model on the triangular lattice, one obtains the following rotated Hamiltonian  
\begin{align}
	\mathcal{H}_{\text{rot}} = J \sum_{\langle i,j\rangle} S_i^y S_j^y + &\cos\left(\theta_i-\theta_j\right)\left(S_i^z S_j^z+S_i^x S_j^x\right) \nonumber \\
	 +&\sin\left(\theta_i-\theta_j\right)\left(S_i^z S_j^x-S_i^x S_j^z\right) \; ,
\end{align}
where $\theta_i$ is $\theta_\text{A}=0$, $\theta_\text{B}=2\pi/3$, $\theta_\text{C}=4\pi/3$ depending on wether the site $i$ belongs to sublattice A, B or C.
Again, we introduce the parameter $\lambda$ in front of all terms which are not diagonal with respect to the rotated ferromagnetic reference state. The final Hamiltonian is then given by
\begin{align}
\label{Eq:XXZ_triangular}
	\mathcal{H}_{\text{rot}}^{\rm XXZ} = J \sum_{\langle i,j\rangle}  - \frac12 S_i^z S_j^z &+ \lambda \left(S_i^y S_j^y - \frac12 S_i^x S_j^x \right. \nonumber \\
		&+\left. \frac{\sqrt{3}}{2} S_i^z S_j^x -\frac{\sqrt{3}}{2} S_i^x S_j^z \right)\; ,
\end{align}
which is of the desired form Eq.~\eqref{Eq::setting}. The order of $i$ and $j$ in the above equation is chosen such that the term $S_i^z S_j^x$ has a positive prefactor. As a reference, we compare our results to the values from Green's function quantum Monte Carlo (GFQMC)~\cite{Capriotti99} and SE~\cite{Zheng06}.

\begin{figure}
\begin{center}
\includegraphics[width=\columnwidth]{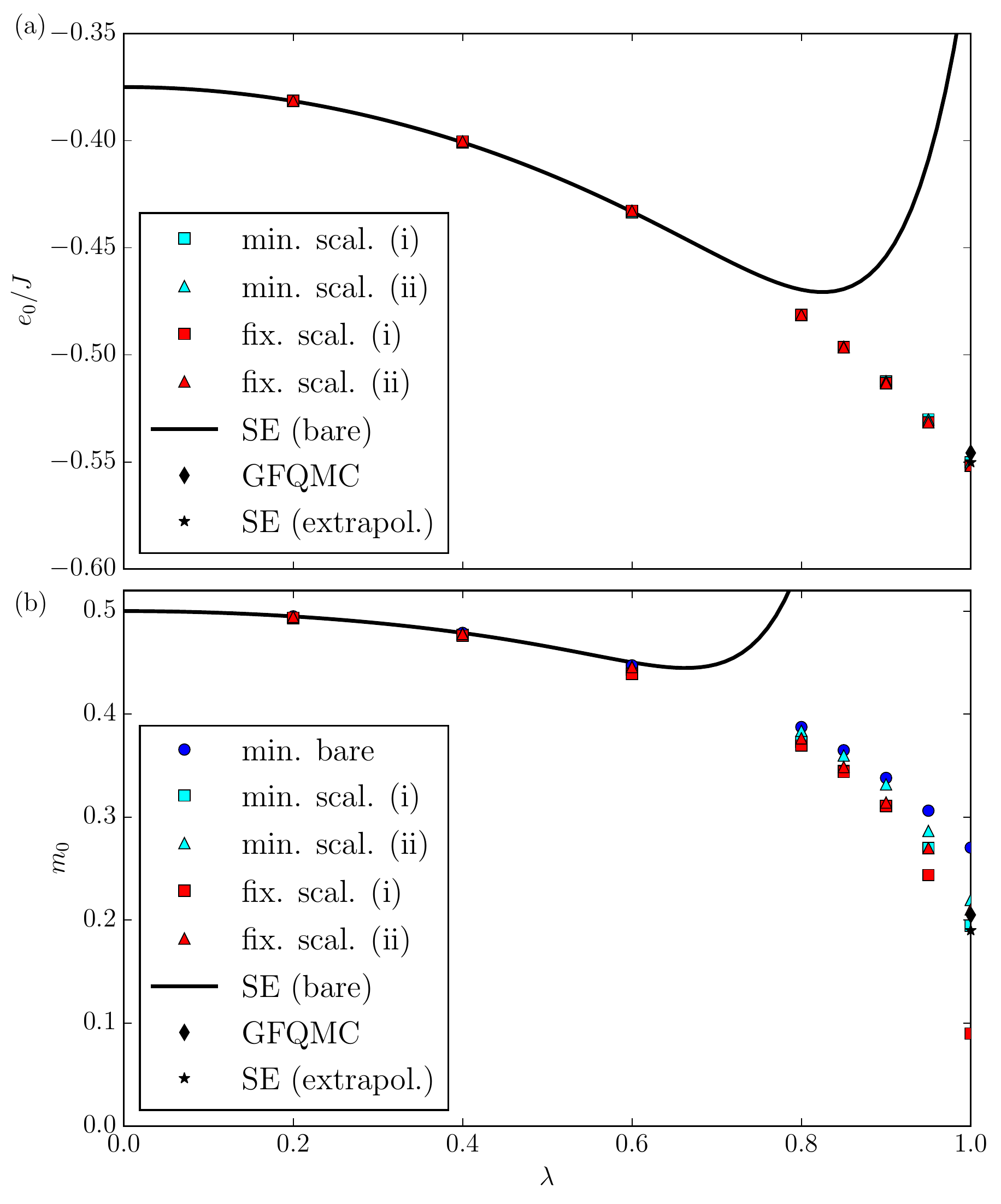}
\end{center}
\caption{(Color online) (a) Ground-state energy per site $e_0/J$ and (b) sublattice magnetization $m_0$ for the XXZ-model on the triangular lattice obtained by scalings of type (i) (squares) and type (ii) (triangles) using the square-graph expansion and the minimum- (cyan symbols) as well as the fixpoint-method (red symbols). Blue circles are the bare results of order 5 and the solid line is the bare order-$13$ SE result~\cite{Zheng06}. At the Heisenberg point $\lambda=1$, the star depicts the extrapolated SE~\cite{Zheng06} whereas the diamond corresponds to the GFQMC-value~\cite{Capriotti99}.}
\label{fig:NLCE_triangularOneHalf_XXZ}
\end{figure}

We start with the discussion of the ground-state energy per site for the Heisenberg case $\lambda=1$. The values from GFQMC and extrapolated SE are $e_0^{\rm GFQMC} = −0.5458(1) J$ and $e_0^{\text{SE}}=-0.5502(4) J$. Our edge-field NLCE data using the arithmetic and the square-graph expansion are displayed in Fig.~\ref{fig:NLCE_triangularOneHalf_arithmeticSquare}. Similarly to the spin-$1/2$ square lattice Heisenberg model, the edge-field is again crucial to get a meaningful NLCE for both expansions. In the square-graph expansion, all NLCE orders display a well-defined minimum as a function of $K$. Notably, these minima converge very well and, in contrast to the square lattice case, monotonously with the NLCE order. Already the bare NLCE results are rather good. The minimal value of the order-$5$ square-graph expansion is \mbox{$e_0^{\text{min}}=-0.548258 J$} which differs only by about $0.002 J$ from $e_0^{\text{SE}}$ and $e_0^{\rm GFQMC}$. This is different for the arithmetic expansion where the bare value of the highest order is at slightly higher values and the convergence with increasing order is less compared to the square-graph expansion. Interestingly, the bare order-$5$ value from the square-graph expansion is already below the GFQMC value which suggests that the extrapolated value from SE is likely to be trusted more than the one from GFQMC, since the values from the edge-field NLCE monotonously decrease with the NLCE order for the square-graph expansion. These conclusions are further strengthened when performing scalings of type (i) and (ii) as shown in the insets of Fig.~\ref{fig:NLCE_triangularOneHalf_arithmeticSquare} and Tab.~\ref{tab:triangularEnergy}. For the arithmetic expansion, both scalings yield different values. While type (i) is very close to the value from SE, the scaling of type (ii) gives a considerably too low value of the energy. In contrast, for the square-graph expansion, both scalings work very well and yield a value within the error estimation of the SE extrapolation~\cite{Zheng06}. 

Next we turn to the sublattice magnetization $m_0$ of the $120^\circ$ long-range ordered ground state of the triangular lattice Heisenberg model. The ordered moment obtained from SE and GFQMC is $m_0^{\text{SE}}=0.19$ and \mbox{$m_0^{\text{GFQMC}}=0.205(10)$} which is lower than in the square lattice Heisenberg model due to the geometric frustration. As for the ground-state energy per site, the edge-fields are essential to regularize the NLCE. The arithmetic and the square-graph NLCE display well defined minima as a function of $K/J$ (see Figs.~\ref{fig:NLCE_triangularOneHalf_arithmeticSquare} and~\ref{fig:NLCE_triangularOneHalf_arithmeticSquare_mag}). These minima decrease monotonously towards the values from SE and GFQMC, but, as for the square lattice Heisenberg model, the bare edge-field NLCE sublattice magnetization is not as close to SE and GFQMC as the ground-state energy per site, e.g.~the bare order-$5$ NLCE minimum is $m_0^{\text{min}}\approx 0.27$ for the square-graph expansion. Again, we have performed scalings of type (i) and (ii) shown in the insets of Fig.~\ref{fig:NLCE_triangularOneHalf_arithmeticSquare_mag}. It can be clearly observed that the quality of both scalings is rather poor for the arithmetic expansion. We therefore show also a linear fit in $1/\mathcal{L}$ which gives a reasonable value of the sublattice magnetization in the thermodynamic limit. In contrast, the scaling for the square-graph expansion works very well yielding values close to the ones from SE and GFQMC. The only exception is the scaling of type (i) for the fixpoint method which reflects the sensitivity due to the limited number of data points.  

In our opinion the different quality of the arithmetic and square-graph expansion can be understood as follows. In the arithmetic expansion a given NLCE order contains several graphs with very similar typical length scales $\mathcal{L}_{\rm arith}$, but with different aspect ratios $L_x/L_y$. It is therefore reasonable that a scaling in $1/\mathcal{L}_{\rm arith}$ is complicated. This is different for the square-graph expansion. Here each NLCE order is clearly dominated by the defining length scale $\mathcal{L}_{\rm sq}$ of the largest quadratic cluster and one therefore expects a better scaling behavior in this length scale. The same reasoning holds also for the square lattice Heisenberg model, but here these properties are likely not observed due to the annoying even-odd effect which is absent on the triangular lattice.

\subsection{Spin-$1/2$ XXZ-model on the triangular lattice}

Next we discuss the full $\lambda$-axis between 0 and 1 in the XXZ-model on the triangular lattice as given by Eq.~\eqref{Eq:XXZ_triangular} after the appropriate sublattice rotation corresponding to the 120$^\circ$ order. As for the same model on the square lattice, one expects that the edge-field NLCE converges better when decreasing $\lambda$ from 1 to 0, since a gap opens for $\lambda<1$ introducing a finite correlation length. This is indeed the case as illustrated in Fig.~\ref{fig:NLCE_triangularOneHalf_XXZ} showing the results for the square-graph expansion.

For small values of $\lambda$ the edge-field NLCE is indistinguishable to the perturbative high-order SE. In contrast to the unfrustrated XXZ-model on the square lattice, the series is alternating in $\lambda$ and it becomes therefore problematic for $\lambda>0.6$. The SE can be improved by extrapolating the alternating series with Pad\'e  extrapolants. The bare order-5 edge-field NLCE behaves smoothly on the full $\lambda$-axis and captures already well the global behavior. Scalings of the bare data become important for $\lambda>0.6$ which corresponds to the same large $\lambda$-regime where the bare SE is unreliable. Altogether, the edge-field NLCE with square graphs captures the physics of the XXZ-model well. It is only the scaling of the sublattice magnetization which becomes challenging close to $\lambda=1$.
We want to note, that we restricted $b_1$ and $\bar{b}_1$ in the scalings to positive values for $\lambda<1$.

\subsection{Spin-$1$ triangular lattice Heisenberg model}\label{Sect:results:triangularOne}

\begin{figure}
\begin{center}
\includegraphics[width=\columnwidth]{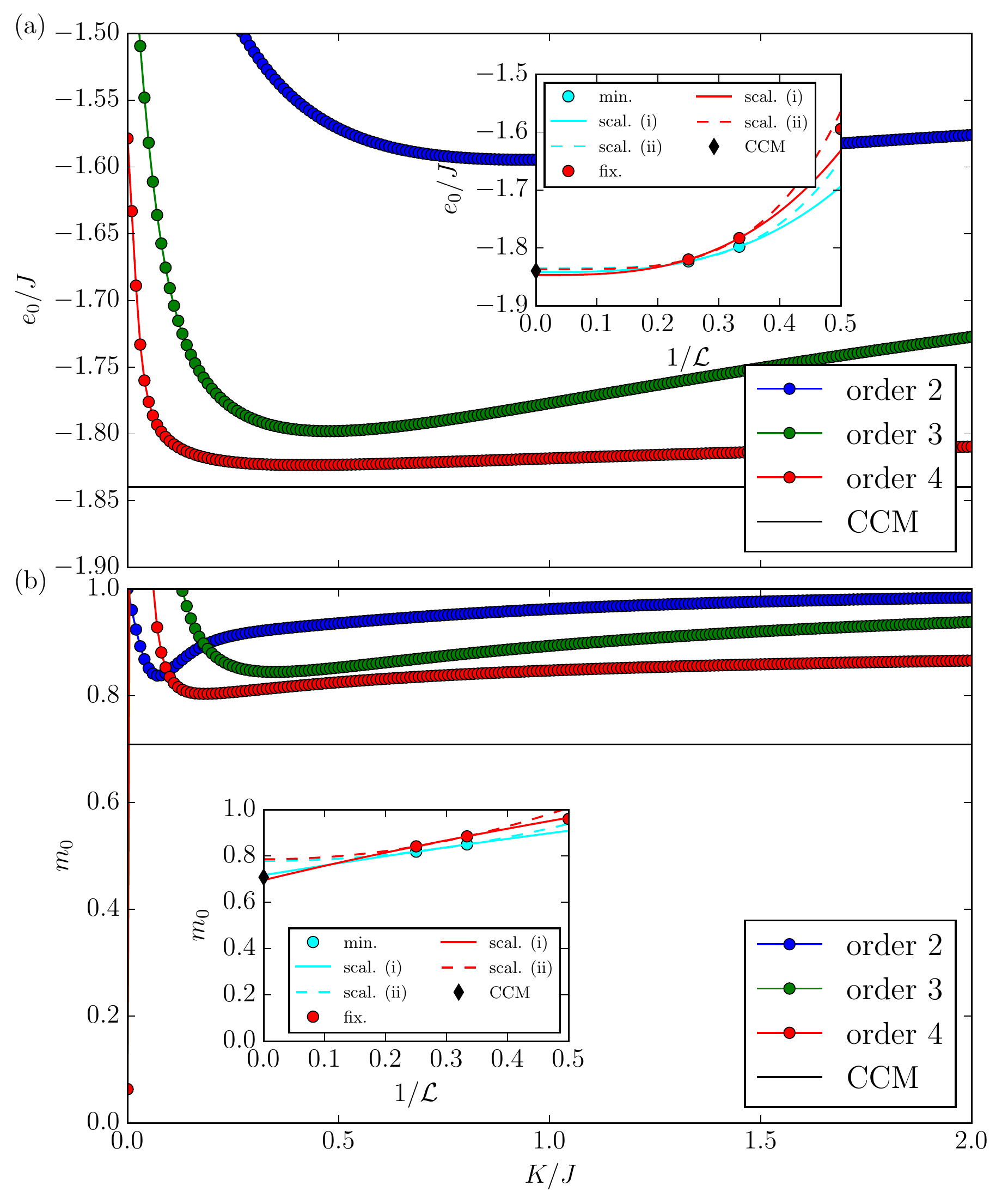}
\end{center}
\caption{(Color online) (a) Ground-state energy per site $e_0/J$ and (b) sublattice magnetization $m_0$ for the spin-$1$ triangular lattice using the square-graph expansion. Different colored symbols correspond to differend orders of the graph expansion. The lines between them are guide to the eyes. Insets show scalings of type (i) (solid lines) and type (ii) (dashed lines) as a function of $1/\mathcal{L}$ through the minimal (cyan circles) and fixpoint (red circles) NLCE values. The black diamond depicts the CCM value from Ref.~\onlinecite{Li11}.}
\label{fig:NLCE_triangularOne_squareExpansion}
\end{figure}

In this subsection we discuss our edge-field NLCE results using square graphs for the spin-$1$ triangular lattice Heisenberg model shown in Fig.~\ref{fig:NLCE_triangularOne_squareExpansion} and Tab.~\ref{tab:triangularEnergy}.  As a reference, we compare to the CCM data from Ref.~\onlinecite{Li11} which are $e_0^{\text{CCM}}=-1.83968 J$ for the ground-state energy per site and $m_0^\text{CCM}=0.7086$ for the sublattice magnetization. 

The anticipated trend from the spin-$1/2$ Heisenberg model on the triangular lattice is clearly visible. The edge-field NLCE displays well-defined minima which approach the values from the CCM monotoneously with increasing NLCE order. The only exception is the order-2 curve for the sublattice magnetization which we attribute to the low order as in the square lattice case. Unfortunately, as for the spin-$1$ square lattice Heisenberg model, the maximal NLCE order is reduced to four compared to the spin-$1/2$ case due to the larger Hilbert space. We also observe that the bare order-4 NLCE is not as close to the CCM as for the unfrustrated square lattice case. The reason might be that the square-graph NLCE does not converge so well due to the geometric frustration, which, however, is also present in the spin-$1/2$ case where quantum fluctuations are even stronger and the geometric frustration is also present. 

The scaled values, along with their deviation from the value obtained by series expansion, are shown in Tab.~\ref{tab:triangularEnergy}.
We remark that, as for the spin-$1/2$ case on the triangular lattice, the scaling of type (i) performs better for the sublattice magnetization.

\setlength{\tabcolsep}{10pt}
\begin{table*}[ht]
\begin{tabular}{llllll}
&&\textbf{Spin-$1/2$ triangular lattice}&&&\\[4pt]
Method 			& Extrapolation 					& $e_0/J$ 		& $|e_0-e_0^\text{SE}|/J$	& $m_0$ 		& $m_0-m_0^\text{SE}$ \\ \hline\\[-6pt]
arith. (fix.)	&	scaling (i) ($4$, $6$, $8$)		& $-0.55301$	& $0.0028$				& --			& -- \\[4pt]
arith. (fix.)	&	scaling (ii) ($4$, $6$, $8$)	& $-0.54778$	& $0.0024$				& $0.0438$		& $0.15$ \\[4pt]
square (fix.)	&	scaling (i) ($3$-$5$)			& $-0.55178$	& $0.0016$				& $0.0898$		& $0.10$ \\[4pt]
square (fix.)	&	scaling (ii) ($3$-$5$)			& $-0.55110$	& $0.00090$				& $0.2104$		& $0.020$ \\[4pt]
arith. (min.)	&	scaling (i) ($4$, $6$, $8$)		& $-0.55451$	& $0.0043$				& --			& -- \\[4pt]
arith. (min.)	&	scaling (ii) ($4$, $6$, $8$)	& $-0.54877$	& $0.0014$				& $0.0646$		& $0.13$ \\[4pt]
square (min.)	&	scaling (i) ($3$-$5$)			& $-0.55015$	& $0.00005$				& $0.1949$		& $0.0049$ \\[4pt]
square (min.)	&	scaling (ii) ($3$-$5$)			& $-0.55049$	& $0.00029$				& $0.2196$		& $0.030$ \\[4pt]\hline\hline\\

&&\textbf{Spin-$1$ triangular lattice}&&&\\[4pt]
Method 			& Extrapolation 						& $e_0/J$ 		& $|e_0-e_0^\text{CCM}|/J$	& $m_0$ 		& $m_0-m_0^\text{CCM}$ \\ \hline\\[-6pt]
square (fix.)	&	scaling (i) ($3$, $4$)				& $-1.84689$	& $0.0072$					& $0.6961$		& $0.012$ \\[4pt]
square (fix.)	&	scaling (ii) ($3$, $4$)				& $-1.83704$	& $0.0026$					& $0.7862$		& $0.078$ \\[4pt]
square (min.)	&	scaling (i) ($3$, $4$)				& $-1.84180$	& $0.0021$					& $0.7175$		& $0.0089$ \\[4pt]
square (min.)	&	scaling (ii) ($3$, $4$)				& $-1.83502$	& $0.0047$					& $0.7790$		& $0.070$ \\[4pt]\hline\hline
\end{tabular}
\caption{Comparison of ground-state energies per site $e_0/J$ and sublattice magnetizations $m_0$ on the spin-$1/2$ and spin-$1$ triangular lattice. The determination method for $K$ is denoted in brackets after the used graph-expansion method (min. for the minimum method and fix. for the fixpoint method). The used orders for the extrapolation are stated after the extrapolation method.}
\label{tab:triangularEnergy}
\end{table*}

\subsection{Spin-$1$ kagome Heisenberg model}
\label{Sect:spin1kagome}
In contrast to all other cases discussed so far, which were long-range spin-ordered states with a finite sublattice magnetization, here we apply the edge-field NLCE to the trimerized ground state of the spin-$1$ Heisenberg model on the kagome lattice \cite{Changlani15,Liu15,Li15,Ghosh15,Oitmaa15,Ixert15}. In the trimerized ground state the symmetry between up and down triangles on the kagome lattice is sponteaously broken. It is therefore possible to adiabatically connect the trimerized ground state to the limit of isolated up (or down) triangles where the ground state is given by the exact product state of singlets on the up (or down) triangles. The elementary excitations above the trimerized ground state of the kagome Heisenberg model have a finite gap \cite{Changlani15} and it is therefore expected that the NLCE converges better compared to the gapless Heisenberg points discussed above on the square and triangular lattice.

We introduce the real parameter $\lambda\in\left[0,1\right]$ which allows to interpolate between the limit of isolated up-triangles $\lambda=0$ and the isotropic Heisenberg model at $\lambda=1$. The Hamiltonian is then defined as
\begin{equation}
\mathcal{H} = J \sum_{\langle i,j\rangle\in\Delta } {\mathbf S}_{i}\cdot{\mathbf S}_{j} +  \lambda J \sum_{\langle i,j\rangle\in\nabla } {\mathbf S}_{i}\cdot{\mathbf S}_{j}\; , 
\label{eq:ham}
\end{equation}
so that the first (second) sum runs over all nearest neighbor sites on up-triangles (down-triangles). In the following we focus on an antiferromagnetic exchange constant $J>0$.

We have performed the edge-field NLCE up to order six using a full-graph expansion in terms of elementary up-triangles. The NLCE order of a graph is defined as the number of triangles $N_{\rm tr}$ of this graph. As we have shown in Ref.~\onlinecite{Ixert15}, the NLCE without edge-fields shows an almost erratic behavior for increasing NLCE order. This behavior can be traced back to a different quantum critical behavior of the one-dimensional unfrustrated chain graphs (in terms of triangles) occuring at values $\lambda\approx 0.8$ well before the isotropic point $\lambda=1$. As a consequence, the reduced contributions of these chain graphs do decay only algebraically with the NLCE order for $\lambda\approx 0.8$ while the embedding factor grows exponentially with the NLCE order leading to a partially diverging NLCE data sequence. Let us note that in Ref.~\onlinecite{Ixert15} we also formulated a reorganized expansion in terms of highly symmetric clusters which gets rid of this issue. Here we will show that the edge-field NLCE also removes the diverging subseries for the full-graph expansion and leads to even better results.

\begin{figure}
\begin{center}
\includegraphics[width=\columnwidth]{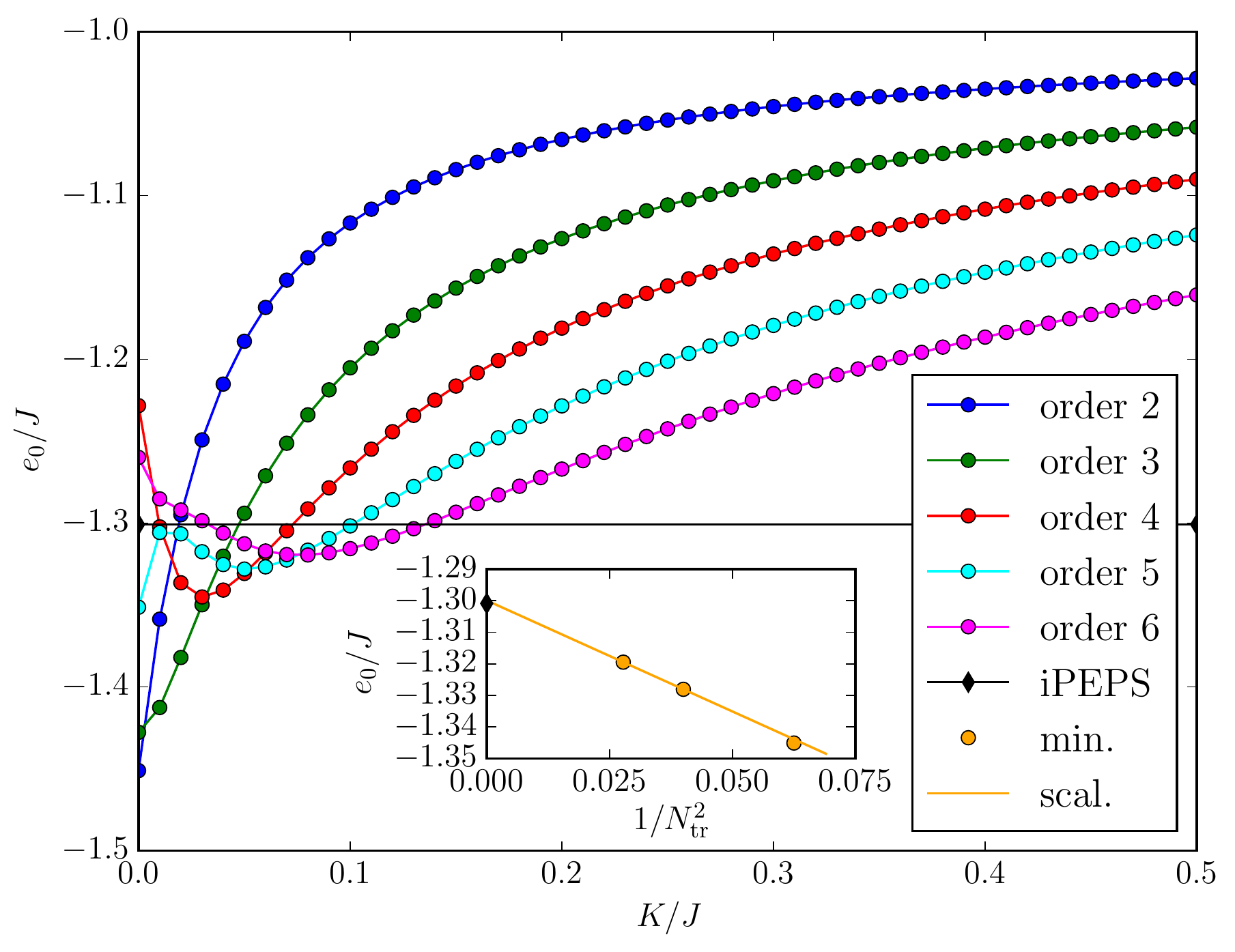}
\end{center}
\caption{(Color online) Ground-state energy per site $e_0/J$ for the spin-$1$ kagome lattice and $\lambda=0.82$. Different colored symbols correspond to different orders of the graph expansion. The lines between them are guide to the eyes. \textit{Inset:} Linear scaling (orange line) through the minimal values (orange circles) of order $5$ and $6$ as a function of $1/N_{\rm tr}^2$. The black diamond depicts the iPEPS result from Ref.~\onlinecite{Liu15}.}
\label{fig:NLCE_kagomeOne_lambda0_82}
\end{figure}
\begin{figure}
\begin{center}
\includegraphics[width=\columnwidth]{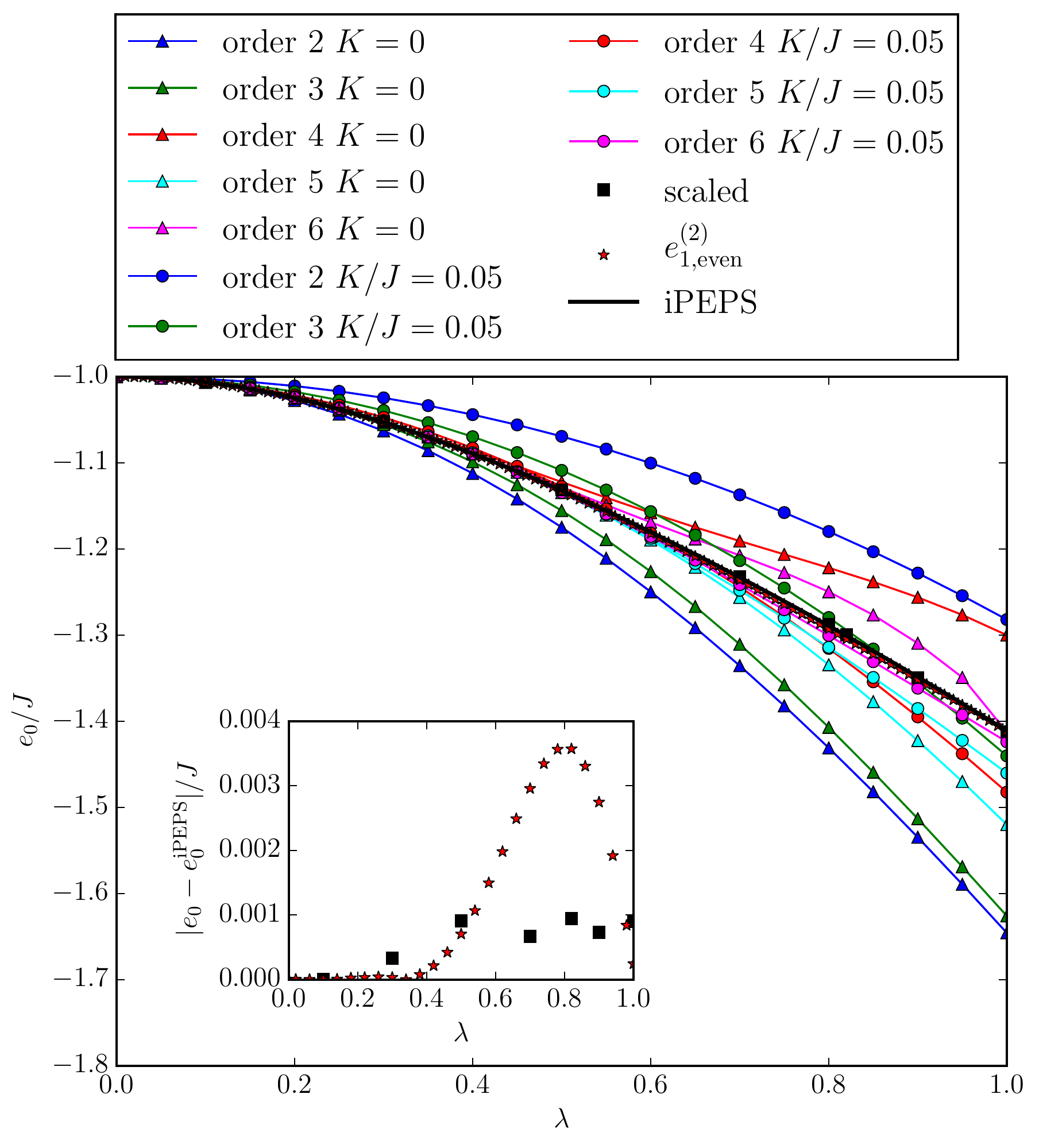}
\end{center}
\caption{(Color online) Ground-state energy per site $e_0/J$ for the spin-$1$ kagome lattice as a function of $\lambda$. Triangles are the results without edge-fields from Ref.~\onlinecite{Ixert15}, whereas circles are results with a small edge-field. The corresponding lines are guide to the eyes. Black squares depict the scaled values of order $5$ and $6$, red stars are the best results of the reorganized graph expansion from Ref.~\onlinecite{Ixert15} and the solid black line illustrates the iPEPS result with bond dimension $D^{*}=5$ from Ref.~\onlinecite{Liu15}. \textit{Inset:} Difference between the ground-state energy per site from the scaled edge-field NLCE or the reorganized expansion from Ref.~\onlinecite{Ixert15} and the ground-state energy per site from iPEPS \cite{Liu15}.}
\label{fig:NLCE_kagomeOne}
\end{figure}

We begin our discussion with the specific case $\lambda=0.82$. This value is close to the quantum critical point of the spin-$1$ triangle chain \cite{Ixert15} and should therefore be most problematic for the full-graph expansion. Furthermore, we can compare our NLCE data directly with the iPEPS-value from Ref.~\onlinecite{Liu15}. The NLCE ground-state energy per site for $\lambda=0.82$ is shown in Fig.~\ref{fig:NLCE_kagomeOne_lambda0_82} as a function of $K/J$ for all NLCE orders up to six. Clearly, the NLCE without (or very small) edge-fields varies strongly for different NLCE orders. In contrast, the NLCE orders 4 to 6 display well-defined minima in $K/J$ which approach monotonously the iPEPS value from below. Although there is no obvious length scale, which can be well defined in the full-graph expansion, we observe heuristically that our NLCE data scales almost linearly in $1/N_{\rm tr}^2$. A linear fit through orders $5$ and $6$ in $1/N_{\rm tr}^2$ leads to a scaled value which is in very good agreement with the iPEPS-value, as shown in the inset of Fig.~\ref{fig:NLCE_kagomeOne_lambda0_82}. The edge-fields are therefore well suited to regularize the NLCE in this most problematic $\lambda$-regime. The same is true on the full $\lambda$-axis. This can be already seen in Fig.~\ref{fig:NLCE_kagomeOne} where we compare the NLCE without edge-fields to the case of a finite field $K/J=0.05$. It is clearly visible that the NLCE data without edge-fields are erratic (especially around $\lambda\approx 0.8$) while the edge-field smoothens the NLCE for all values of $\lambda$. 

As for $\lambda=0.82$, we have performed scalings for various values of $\lambda$ shown as black squares in Fig.~\ref{fig:NLCE_kagomeOne}. Remarkably, they agree very well with the iPEPS-data. The differences between the energies from scaled edge-field NLCE and iPEPS is even smaller than the same difference using the reorganized NLCE from Ref.~\onlinecite{Ixert15} as can be seen in the inset of Fig.~\ref{fig:NLCE_kagomeOne}. Interestingly, the reorganized NLCE displays the largest difference to the iPEPS energy for $\lambda\approx 0.8$, i.e.~it still ``feels'' the instability observed in the full-graph expansion without edge-fields (although on a much smaller energy scale). In contrast, the edge-field NLCE shows an almost constant difference to the iPEPS values with constant bond dimension $D^{*}=5$ for $\lambda\in [0.5,1.0]$ which we attribute mostly to uncertainties in the scaling (as well as to the finite $D^{*}$). For $\lambda=1$ -- corresponding to the isotropic Heisenberg model on the kagome lattice -- our best result from Ref.~\onlinecite{Ixert15} using a reorganized graph expansion was $e_{1,\text{even}}^{(2)}=1.4114 J$. The full-graph expansion with edge-fields yields the scaled value $e_0^{\text{scaled}}=-1.41203 J$ which is nearly within the error estimations of the value $e_0^{\text{iPEPS}}=-1.4116(4)J$ obtained by the iPEPS calculations.
Furthermore it also compares well to the DMRG-Value $e_0^{\text{DMRG}}=−1.410(2)J$ from Ref.~\onlinecite{Changlani15}.

\section{Conclusions}
\label{Sect:conclusions}
In this work we have presented a generic scheme to perform NLCEs in long-range ordered quantum phases. The essential idea is to incorporate the effect of the long-range order into the exact diagonalization on graphs. Then each graph ``remembers'' the ordered quantum phase out of which it is taken from and the correct quantum fluctuations are captured in the NLCE. This is achieved by assuming an appropriate ordered reference state outside each graph so that the edge-couplings give rise to an effective edge-field. The edge-field breaks the relevant symmetry, which corresponds to the underlying long-range order, and it ensures that the NLCE is carried out for each graph inside the correct quantum phase. The field is subextensive, since it scales with the perimeter of the graphs. As a consequence, the NLCE is regularized and fluctuations from different quantum phases and different quantum critical behavior on lower-dimensional subgraphs are excluded.  

We have implemented four different NLCEs. The full-graph expansion as well as the rectangular-graph expansion are complicated to extrapolate to the infinite-order limit, since in the full-graph expansion no proper length scale can be defined while the NLCE data squences are not smooth for different NLCE orders in the rectangular-graph expansion. Notably, we found heuristically for the full-graph expansion in the trimerized ground state of the spin-$1$ Heisenberg model on the kagome lattice, that the NLCE data sequence of the ground-state energy is approximately linear in $1/N^2$. This suggests that scalings might be also valuable for NLCEs using a full-graph expansion. In contrast, both the arithmetic as well as the square-graph expansion give satisfactory results and scalings towards the infinite-order limit can be performed successfully. However, we have observed that the scaling is more complicated in the arithmetic expansion which can be traced back to the fact that different clusters of the same NLCE order have almost the same typical length scale but different aspect ratios. This is different for the square-graph expansion where each order is clearly dominated by the length scale of the largest quadratic cluster. In this work we have also investigated two different scaling laws for the NLCE data. Although the quality of both scaling types is very similar (especially for the ground-state energy per site). We are strongly convinced that it is an important but challenging task to formulate proper scaling laws for NLCEs from first principles.  

We have applied the edge-field NLCE to the long-range spin-ordered phases of the spin-1/2 and spin-1 Heisenberg model on the square and triangular lattice as well as to the trimerized valence bond crystal of the spin-1 Heisenberg model on the kagome lattice. In all cases we have found that edge-fields are essential to obtain a well-behaved NLCE for the ground-state energy per site and the sublattice magnetization. We have chosen the unfrustrated spin-1/2 Heisenberg model on the square lattice to test four different cluster expansions and we formulated two strategies to determine the optimal value of the edge-field. It is found that all NLCEs compare reasonably well with state-of-the-art numerical QMC data. Furthermore, the arithmetic and the square-graph expansion perform best and both approaches to determine the edge-field yield similarly good results. We find it interesting that the square-graph expansion works, in contrast to the arithmetic expansion, even better for the geometrically frustrated spin-1/2 triangular lattice compared to the unfrustrated square lattice in the sense that no even-odd effect is present and therefore scalings can be performed through more NLCE data points. It would be nice if one could formulate a reorganized NLCE with different elementary clusters which does not show this even-odd effect on the square lattice. Overall, we are convinced that the edge-field NLCE can be extended successfully into several directions which we hope to tackle in future works. First, it would be interesting to apply other numerical tools to extract quantitatively physical quantities for larger open clusters. This would allow to improve the quality of the infinite-order scaling considerably resulting in even better NLCE results. Second, one should apply the edge-field NLCE to systems where several quantum phases compete with each other. A comparison of different NLCE energies and their convergence behavior should allow the determination of the ground-state phase diagram. Third, one might speculate whether a formulation of edge-field NLCEs can be achieved where the mean-field reference product state is replaced by more complex entangled tensor networks. This would open the fascinating perspective to tackle also exotic quantum disordered phases with NLCEs.

\begin{acknowledgments}
We thank W. Li for providing numerical data. This research has been supported by the Virtual Institute VI-521 of the Helmholtz association.
\end{acknowledgments}

\end{document}